# Effect of Point Defects and Lattice Distortions on the Structural, Electronic, and Magnetic properties of $Co_2MnAl$ Heusler alloy

Amar Kumar[1], Sujeet Chaudhary[1] and Sharat Chandra[2]

[1]Thin Film Laboratory, Department of Physics, Indian Institute of Technology Delhi, New Delhi 110016, India

[2]Material Science Group, Indira Gandhi Centre for Atomic Research, HBNI, Kalpakkam, Tamil Nadu-603102, India

Corresponding Authors: sujeetc@iitd.ac.in and sharat@igcar.gov.in

## Abstract

The effects of various point defects and lattice distortions on the structural, electronic, and magnetic properties of $Co_2MnAl$ alloy are investigated using pseudopotential plane wave-based density functional theory calculations. For the point defects, six types of binary antisites, three types of ternary antisites, and three kinds of vacancies have been simulated with different disorder degrees, up to a maximum disorder degree of 12.50%. For the lattice distortions, cubic strain within $-10\% \leq \Delta V/V_0 \leq 10\%$ (corresponding to the lattice parameter ranging from 5.50 Å to 5.88 Å) and tetragonal deformed structures with $0.5 \leq c/a \leq 1.5$ at three different unit-cell volumes - $V_0$ and ($V_0 \pm 5\% V_0$) have been considered. The relative formation energies, spin polarization, and magnetic moments for all defected structures are discussed in detail. The Mn-deficient structures resulting from $Al_{Mn}$ binary antisite disorder ($Co_2Mn_{0.9375}Al_{1.0625}$ and $Co_2Mn_{0.875}Al_{1.125}$) and ($Co_{Mn}$+$Al_{Mn}$) ternary antisite disorder ($Co_{2.0625}Mn_{0.875}Al_{1.0625}$) may spontaneously formed during the synthesis of $Co_2MnAl$, due to their negative relative formation energies. In the case of $Mn_{Co}$ binary antisite disordered structures ($Co_{1.9375}Mn_{1.0625}Al$ and $Co_{1.875}Mn_{1.125}Al$) and ($Mn_{Co}$+$Al_{Co}$) ternary antisite disordered structure ($Co_{1.875}Mn_{1.0625}Al_{1.0625}$), the impurity Mn atoms exhibit antiparallel coupling with neighboring Co and Mn atoms, leading to the emergence of half-ferrimagnetism phenomenon. The $Co_{Al}$ and $Mn_{Al}$ binary antisite disordered structures (namely, $Co_{2.0625}MnAl_{0.9375}$, $Co_{2.125}MnAl_{0.875}$, $Co_2Mn_{1.0625}Al_{0.9375}$ and $Co_2Mn_{1.125}Al_{0.875}$) and ($Co_{Al}$+$Mn_{Al}$) ternary antisite disordered structure ($Co_{2.0625}Mn_{1.0625}Al_{0.875}$) exhibit perfect half-metallicity (*c.f.* 100% spin polarization and integer magnetic moment according to the Slater Pauling rule). The rest of the antisite disorders have a marginal effect on the half-metallic properties of Co2MnAl. The disordered structures maintain the high spin polarization ($\geq$ 70%) and nearly same magnetization as that in the *$L2_1$*-ordered structure. Conversely, the vacancy defects significantly affect the electronic and magnetic properties. They are found to cause the narrowing of the pseudogap (Al-vacancies), including vanishing in some cases (Co- and Mn-vacancies). For the lattice distortions, the uniform strain has a minimal effect on spin polarization and magnetization. The cubic strained structures exhibit high spin polarization and constant magnetization. Under negative strain within $-10\% \leq \Delta V/V_0 \leq 7\%$ (for 5.50 Å ≤ $a$ ≤ 5.58 Å), the strained structures have perfect half-metallicity. With increasing strain values, spin polarization decreases monotonically, but still, it is ~60% at $\Delta V/V_0 = 10\%$ ($a$ = 5.88 Å). On the other hand, tetragonal distortions lead to significant degradation in half-metallic behaviour, except for small distortion values ($\Delta c/a$), irrespective to their volume. This degradation results from the considerable change in density of states near Fermi level. This renders the tetragonal distortion undesirable for device applications.

# 1. Introduction

Materials with high spin polarization ($P$) and high Curie temperature ($T_C$) are promising materials for spintronic applications. Full Heusler alloys (HAs) containing Co and Mn as transition metal elements, such as $Co_2MnZ$ (Z is sp-group element) have been theoretically predicted and experimentally demonstrated to be attractive candidates for spintronic applications due to their remarkable properties, specifically, very high spin polarization (~100%, theoretically), high Curie temperature, low saturation magnetization, and structural compatibility with conventional wide-gap semiconductors. In $Co_2MnZ$ HAs, the presence of Co and Mn atoms ensures high $T_C$ and high $P$ at the Fermi level, respectively [1,2]. Although there are many reports for 100% spin polarization in HAs with perfectly ordered structures from the band structure calculations, most of the measurements of spin polarization in HAs yield only modest values, typically ranging from 30% to 50%; which leads to the poor performance of spintronic devices. However, the recent observation of nearly 100% spin polarization in $L2_1$-ordered epitaxial thin films of $Co_2MnSi$ demonstrates that with controlled growth conditions, even ~100% spin polarization is achievable experimentally within HAs [3]. Numerous experimental and theoretical studies indicate that the structural imperfections originating from various point defects and lattice distortions are the prominent reasons for the experimentally observed low spin polarization of HAs. The experimentally synthesized HA thin films are often prone to various point defects and lattice distortions; for example - atomic swap disorder between Fe and Al atoms in $Co_2FeAl$ alloy [4], Co-vacancy in $Co_2TiSb$ alloy [5], antisite disorder between Mn-Ga sites in $Mn_2NiGa$ alloy [6] and meta-stable tetragonal phase formation in $Mn_2CoAl$ upon $Ar^+$ irradiation [7]. Thus, the presence of point defects and lattice distortions within HAs thin films can be considered a common occurrence.

In general, the point defects in the HAs thin film can be categorized into five major categories: site disorder, antisite disorder, vacancy defects, interstitial defects, and impurity defects. Among these; the site disorder (swapping among different kinds of atoms while maintaining the stoichiometry of the parent alloy), antisite disorder (substitution of one kind of atom with another while keeping the total number of atoms the same as the parent alloy, but with different composition), and vacancy defects (absence of constituting atom) are the most observed point defects in Co-based full HAs. The presence of other point defects like interstitials and impurity defects has rarely been reported in the literature [9]. The presence of these structural imperfections in thin films are primarily dictated by the growth conditions, especially thermodynamic conditions for growth. Many recent studies infer that the presence of various point defects (*i.e.*, antisite disorder and vacancy defects) typically leads to the degradation of the half-metallic properties of HAs. However, a few material-specific defects can also coexist where the half-metallic nature and magnetization of HAs remain nearly the same [10-13]. Thus, understanding the effect of various point defects on the structural, electronic, and magnetic properties of Heusler alloys is crucial for spintronics.

On the other hand, the lattice distortions in HAs generally occur either in a uniform form, in a tetragonal manner, or an octahedral manner. Lattice distortion in HAs thin films can occur due to various factors such as lattice mismatch with the substrate or adjacent buffer layer, thermal treatment of the deposited film, specific deposition technique employed, *etc*. In *most cases*, lattice mismatch with adjacent layers leads to uniform (*i.e.*, isotropic) strain or tetragonal (*i.e.*, anisotropic) distortion in the HA thin film. For spintronic applications, for example, $Co_2MnAl$ (lattice parameter, a = 5.70Å) might be required to be deposited on Si substrate (a = 5.43Å) and/or MgO substrate (a = 4. 25Å), resulting in tetragonal distortion of the HA lattice [14,15]. Furthermore, the thermal treatment process for improving the microstructure of the deposited films during or after growth significantly affects the lattice parameters of the deposited films and *usually* leaves the uniformly expanded lattice (or uniform tensile strain) [16–19]. The thin films prepared by the sputter techniques may have stress of ±1 GPa because of the preparation method alone [20–23]. Such substantial stress *generally* results in uniform tensile or compressive strain in the lattice. These distortions can significantly modify the alloys' structural, electronic, and magnetic properties. For instance, in $Co_2FeAl$ alloy, 100% spin polarization is retained under 0 to +4% uniform strain (defined *w.r.t.* the optimized lattice parameter), while it decreases from 90% to 16% for -1% to -6% negative strains. Also, $Co_2FeAl$ is found to be very sensitive to tetragonal distortions as 100% spin polarization is preserved only under cubic strain only [24]. However, unlike the point defects, which often adversely affect material functionality for spintronics, the lattice deformation in HA thin films can also result in some advantageous phenomena. For instance, such deformations can increase the in-plane magnetic anisotropy due to large mismatches with adjacent layers [16,25]. Additionally, HAs with metastable or stable tetragonal phase can also have some useful phenomenon like - perpendicular magnetic anisotropy (PMA), intrinsic exchange bias, shape memory effect, magnetocaloric effect, and wide-ranging applications for sensors and actuators [26]. Therefore, a comprehensive study of the lattice distortion effects, including compressive- and tensile-uniform strain as well as tetragonal distortion on the electronic and magnetic properties of HAs is also equally important. Additionally, the coexistence of a stable or metastable tetragonal phase alongside the cubic phase in Heusler alloys can provide significant potential and benefits.

Among $Co_2MnZ$ full HAs, $Co_2MnAl$ exhibits very high spin polarization of ~76% at Fermi level ($E_F$), saturation magnetization ($M_s$) ~4.02 $\mu_B$/*f.u.* and high $T_c$ of 720 K, making it a promising candidate for spintronics. Here, '*f.u.*' represents 'formula-unit'. However, despite these promising characteristics, there is limited investigation regarding the effects of point defects and lattice distortions on the structural, electronic, and magnetic properties for $Co_2MnAl$ alloy. Galankis *et al.* [27] have studied the effect of various point defects, such as antisite disorder between Mn-Al sites and Co-Mn site, particularly $Al_{Mn}$, $Mn_{Al}$ and $Mn_{Co}$, and vacancies at Co-, Mn- and Al-sites ($V_{Co}$, $V_{Mn}$ and $V_{Al}$) using DFT calculations. Here, $A_B$ antisite disorder is defined as B-atom substituted by A-atom and $V_X$ denotes vacancy defect at X atomic-site in ideal crystal. Both antisite disorders and vacancy defects lead to off-stoichiometric alloy; while maintaining the total number of atoms constant. They find that the presence of $V_{Co}$, $V_{Mn}$ and $V_{Al}$ lead to the depolarization of $Co_2MnAl$; while the $Mn_{Co}$ has mild effect on the spin polarization. Regarding magnetization, total magnetic moment increases for $Mn_{Al}$ antisite disorder, whereas it decreases for $Al_{Mn}$ and $Mn_{Co}$ antisite disorders. More recently, Feng *et al.* [28] have studied the influence of Mn content on half-metallicity and magnetism of off-stoichiometric $Co_2MnAl$ through DFT calculations. For Mn-rich structure, they considered the $Mn_{Co}$ and $Mn_{Al}$ antisite disorders; and for Mn-poor structure, they considered the $Co_{Mn}$ and $Al_{Mn}$ antisite disorders as well as $V_{Mn}$. Mn-rich samples resulting from $Mn_{Co}$ antisite disorder and Mn-poor samples resulting from $Co_{Mn}$ antisite disorder are expected to be observed experimentally due to their lower relative formation energies. Density of states (DOS) calculations show that the presence of $Co_{Mn}$ antisite disorder and $Mn_{Co}$ antisite disorder lead to decreased and increased $P$, respectively; whereas presence of $Al_{Mn}$ antisite disorder leaves $P$ unchanged. As $Mn_{Co}$ disorder concentration increases in $Co_{2-x}Mn_{1+x}Al$, the disordered alloy exhibits 100% spin polarization from $x = 0.50$ to 0.875 and turns into a spin gapless semiconductor at $x = 1$, which is an exceptionally intriguing phenomenon. However, there is a lack of comprehensive study that includes all possible antisite disorders involving transition metal elements and *sp*-group elements, as well as vacancy defects for various types of atoms. To the best of our knowledge, theoretical investigations of $Co_2MnAl$ have not covered the $Al_{Co}$ antisite disorder ($Co_{2-x}MnAl_{1+x}$) and ternary antisite disorder involving three atoms (*i.e.*, $Co_{2+x+y}Mn_{1-x}Al_{1-y}$, $Co_{2-x}Mn_{1+x+y}Al_{1-y}$ and $Co_{2-x}Mn_{1-y}Al_{1+x+y}$). Notably, ternary antisite disorders are favored in experimental growth and have been observed extensively [29]. Also, the spin polarization and magnetization level in the presence of various vacancy defects are yet to be explored. Regarding the lattice distortions effects in $Co_2MnAl$, Nepal *et al.* have explored the impact of lattice parameter variation on $M_s$ of $Co_2MnAl$ by changing the experimental lattice parameter (5.75 Å) by ±2%. They showed that a uniform expansion of lattice by 2% (lattice parameter 5.86 Å) does not change the total magnetic moment from its ideal value (~4.12 $\mu_B$/*f.u.*). However, when the lattice is uniformly contracted by 2% (lattice parameter 5.63 Å), the total magnetic moment increased to ~ 5.00 $\mu_B$/*f.u.* Under lattice distortion, the major change was observed in the magnetic moments of Mn-atoms; whereas the Co-atom's magnetic moments experience minor modifications [30]. However, a comprehensive investigation of lattice distortion effects on $Co_2MnAl$ is lacking; particularly, the impact of uniform strain and tetragonal distortion remain unexplored.

Interestingly, there are also some reports wherein $Co_2MnAl$ is reported to exhibit exceptional behaviors in the presence of structural imperfections, such as - enhanced spin polarization in the presence of B2 site disorder as compared to the *$L2_1$* ordered structure reported by utilizing *ab initio* DFT calculations [32], emergence of high perpendicular magnetic anisotropy in the presence of lattice distortions and antisite disorders, demonstrated experimentally [31], and spin gapless semiconducting (SGS) nature in the presence of high disorder degree of $Mn_{Co}$ antisite disorder observed experimentally [29] as well as employing DFT calculations [28]. Remarkably, such anomalous behavior of $Co_2MnAl$ in the presence of structural perturbations are rarely observed for other Co-based HAs. Given the partial availability of results for point defects and lattice distortions, such occurrences in presence of structural imperfections further motivate and emphasize the necessity of comprehensively exploring the effects of structural imperfections on the structural, electronic, and magnetic properties of $Co_2MnAl$ alloy vis-a-vis the type of defects. Therefore, we have tried to systematically study the effect of various point defects (binary antisite disorder, ternary antisite disorder, and vacancy defects) and lattice distortions (uniform tensile strain, uniform compressive strain, and tetragonal distortion) on the structural, electronic, and magnetic properties of $Co_2MnAl$ alloy.

The rest of this paper is organized as follows: First, the computational methodology adopted for the present study is briefly discussed in Section 2. Then, the validation of different exchange-correlation functionals to be used for studying the physical properties of the $Co_2MnAl$ alloy is discussed in Section 3.1. This is followed by the presentation of comprehensive results on the effect of the presence of point defects (binary antisite, ternary antisite, and vacancy defects with different disorder concentrations) and lattice distortions (uniform strains and tetragonal distortions) on the structural, electronic, and magnetic properties of $Co_2MnAl$ alloy in Sections 3.2 & 3.3. Finally, a detailed summary of all the findings is provided in the concluding Section in Section 4.

# 2. Computational details

The plane wave pseudopotential-based density function theory calculations, as implemented in QUANTUM ESPRESSO code, are carried out to study the effect of various point defects and lattice distortions on structural, electronic, and magnetic properties of $Co_2MnAl$ Heusler alloys [33,34]. The generalized gradient approximation (GGA) and GGA+U method have been used to deal with the electronic exchange and correlation interactions with 'U' representing the Hubbard parameter within the linear-response method of Cococcioni [35]. The scalar relativistic projector augment wave pseudopotentials from PSlibrary with valence-electrons configurations of Co ($3s^23p^64s^23d^7$), Mn ($3s^23p^64s^23d^5$) and Al ($3s^23p^1$) are used to deal with electron–ion interaction correctly [36]. The validations of exchange-correlation functional for $Co_2MnAl$ alloy and the effect of lattice distortions on the structural, electronic, and magnetic properties of $Co_2MnAl$ have been studied using 16-atoms unit-cell; whereas for study the effect of point defects the structural, electronic and magnetic properties of $Co_2MnAl$, 64 atoms supercell (2×2×1 supercell) have been used. For the 16-atoms unit-cell, the energy converged Monkhorst-Pack 15×15×15 k-point grid and plane-waves cut-off energy of 250 Ry are chosen for calculations. Accordingly, for the 64-atoms supercell, 7×7×15 sized Monkhorst-Pack grid is used to maintain the same k-points density between 16-atoms and 64-atoms unit-cell. This causes a very small energy difference of < 0.1 meV/*f.u.* in total energies between 16-atoms and 64-atoms unit-cells. The optimized tetrahedron method is employed to perform the Brillouin zone integration [37]. The convergence threshold for self-consistency is set to be $10^{-6}$ Ry. For lattice distortions, only atomic positions; whereas for full-cell relaxation, atomic positions along with the cell parameters are allowed to relax to find out the optimized structure, with a force convergence criteria of $10^{-3}$ Ry/bohr. As spin-orbit coupling has minor effect on structural, electronic, and magnetic properties of $Co_2Mn$-based full Heusler alloys, therefore the effect of spin–orbit coupling is not considered for the calculations [2].

# 3. Results and Discussion

## 3.1 Exchange Correlation functional validation for $Co_2MnAl$

Numerous approximations exist for exchange-correlation (XC) functionals, with three primary are- Local Density Approximation (LDA), Generalized Gradient Approximation (GGA) and Hybrid functionals. The accuracy of DFT calculations strongly depends upon the approximation used for XC functional. To ensure that the chosen XC does not influence our findings, we first benchmarked the adequate XC functional for $Co_2MnAl$ alloy. Hybrid functional are well known for accurately describing a wide range of electronic properties like lattice parameters, magnetic moments, and band gaps. However, they required very high computational costs due to the incorporation of non-local potential. Therefore, we have not used Hybrid XC functional for our calculations. GGA XC approximates almost accurately the structural, electronic, and magnetic properties of a metallic system compared to LDA; therefore, first, we have calculated the structural, electronic, and magnetic properties of $Co_2MnAl$ using GGA XC functional.

Also, since $Co_2MnAl$ contains the 3d-transition elements (Co and Mn), accounting for the on-site coulomb repulsion between the localized d-electrons is needed for better predictions of structural, electronic, and magnetic properties. Therefore, to include the on-site coulomb repulsion between the electrons in the d-orbitals; we have incorporated the Hubbard U correction in the DFT calculations with GGA functional. Here, U represents the on-site coulomb potential, which can be understood as an add-on to GGA XC. The U values are calculated using the linear-response method of Cococcioni [35]. The calculated Hubbard parameter for the Co and Mn atoms is 5.93 eV and 4.41 eV respectively, well within the typical range (between 1 eV - 5 eV) observed for most 3d-transition elements. The calculated DOS plots without and with the specified U values are given in supplementary material in Figures S1(a) & S1(b), respectively. These U values are relatively large as compared to the reported values of 2.5 eV – 3.5 eV for Co and Mn atoms in $Co_2MnAl$ [38], but it is important to note that the U values depend upon the calculation method, choice of pseudopotentials and basis-set. Thus, they are not fixed and can have different values with different inputs for the same element. The calculated optimized lattice parameter a (Å), total magnetic moment $M_s$ ($\mu_B$/f.u.), atomic magnetic moment ($m_X$) and spin polarization (*P*) using the GGA and GGA+U method, with the obtained U values, are presented in Table 1 along with other results wherever available. As seen from Table 1, the GGA+U optimized lattice parameter, total magnetic moments and spin polarization significantly exceed the experimental and GGA findings. In contrast, the results obtained using GGA are consistent with the experimental results. Therefore, this study shows that GGA+U method with specified U values is not helpful for $Co_2MnAl$, and GGA is an adequate XC for studying the properties of $Co_2MnAl$. As per our best knowledge, this observation is also consistent with the other available GGA+U studies for $Co_2MnAl$ so far, which further supports the validation of the incompatibilities of the GGA+U method to study $Co_2MnAl$ alloy [32,39,42]. Therefore, GGA XC would be used for further calculations.

**Table 1:** The calculated Hubbard parameters ($U_{Co}$ for Co atoms and $U_{Mn}$ for Mn atoms), optimized lattice parameter *a* (Å), total magnetic moment *Ms* ($\mu_B$/*f.u.*), atomic magnetic moments ($m_{Co}$, $m_{Mn}$, $m_{Al}$ for Co, Mn and Al atomic moments, respectively) and spin polarization (*P*) for $L2_1$ ordered $Co_2MnAl$ with different XC functional. The other theoretical and experimental results are also provided for comparison. Exp. denotes experimental data, and lattice parameters marked with asterisk (*) correspond to the results obtained for lattice parameter fixed at the experimental value (*i.e.,* 5.75 Å) for bulk $Co_2MnAl$ alloy used for calculating the electronic and magnetic properties.

| **Properties** | **Present study (GGA)** | **Present study (GGA+U)** | **Tsirogiannis *et al.*** [39] **(GGA+U)** | **Nepal *et al.*** [30] **(GGA+U)** | **Webester *et al.*** [40] **(Exp.)** | **Guillemard *et. al.*** [41] **(Exp.)** |
|---|---|---|---|---|---|---|
| $U_{Co}$ **(eV)** | - | 5.93 | 2.70 | 2.70 | - | - |
| $U_{Mn}$ **(eV)** | | 4.40 | 3.65 | 3.65 | - | - |
| ***a*** **(Å)** | 5.69 | 6.23 | 5.75* | 5.75* | 5.75 | 5.76 |
| $M_s$ **($\mu_B$/*f.u.*)** | 4.02 | 7.25 | 6.44 | 4.86 | 4.01 | 4.32 |
| $m_{Co}$ **($\mu_B$)** | 0.8 | 1.54 | 1.38 | 0.87 | 0.5 | - |
| $m_{Mn}$ **($\mu_B$)** | 2.59 | 4.29 | 4.18 | 3.43 | 3.01 | - |
| $m_{Al}$ **($\mu_B$)** | -0.07 | - | - | - | - | - |
| ***P*** | 76.16% | 80.71% | - | - | - | 63% |

## 3.2 Effect of point defects on structural, electronic and magnetic properties of $Co_2MnAl$

In this section, the effect of various point defects, namely binary antisite disorders, ternary antisite disorders, and vacancy defects on the structural, electronic, and magnetic properties of $Co_2MnAl$ using 16 *f.u.* unit-cell (*i.e.*, 64 atom supercell) is discussed. For binary antisite disorders; $Co_{Mn}$ antisite ($Co_{2+x}Mn_{1-x}Al$), $Co_{Al}$ antisite ($Co_{2+x}MnAl_{1-x}$); $Mn_{Co}$ antisite ($Co_{2-x}Mn_{1+x}Al$), $Mn_{Al}$ antisite ($Co_2Mn_{1+x}Al_{1-x}$); $Al_{Co}$ antisite ($Co_{2-}$ ${}_{x}$MnAl$_{1+x}$,) and $Al_{Mn}$ antisite ($Co_2Mn_{1-x}Al_{1+x}$) have been taken into consideration. Since in the experiments, atomic vacancy can occur at any atomic site; therefore, the mono- and di-vacancies for Co-atomic site ($Co_{2-x}MnAl$ or $V_{Co}$), Mn-atomic site ($Co_2Mn_{1-x}Al$ or $V_{Mn}$), and Al-atomic site ($Co_2MnAl_{1-x}$ or $V_{Al}$) have been examined. For ternary antisite disorder; ($Mn_{Co}$+$Al_{Co}$)-antisite disorder ($Co_{2-2x}Mn_{1+x}Al_{1+x}$ or Co-deficient structure), ($Co_{Mn}$+$Al_{Mn}$)-antisite disorder ($Co_{2+x}Mn_{1-2x}Al_{1+x}$ or Mn-deficient structure), and ($Co_{Al}$+$Mn_{Al}$)-antisite disorder ($Co_{2+x}Mn_{1+x}Al_{1-2x}$ or Al-deficient structure) have been considered. The value of $x$, which is directly connected to the disorder concentration through the number of substituted atoms, is taken as 0.0625 and 0.125 for modeling disordered structures for binary antisite disorders and vacancy defects; whereas the ternary antisite disorders have been modeled using $x$=0.0625.

The value of $x$ is chosen considering the following facts: As reported in the literature, high disorder concentration for the antisite disorders and vacancy defects result in very high positive or negative relative formation energies for the defected structure. For the significantly high positive relative formation energies, defects are easily removable *via* high-temperature processing, *viz.*, high-temperature thin film growth and/or post-deposition annealing. Conversely, exceedingly negative relative formation energies at high concentrations can lead to phase separation [43,44]. Additionally, detecting very small disorder concentrations using conventional experimental techniques presents a challenge [45]. For this reason, in most experiments where the antisite disorders and vacancy defects have been reported, the disorders concentration range found varies from 5% to 15%. For example - around 5% of the Co sites are occupied by Mn atoms and 14% of the Mn sites are occupied by Co atoms in $Co_2MnSi$ alloy [46]; 10% antisite disorder between Ti-Al sites in $Co_2TiAl$ alloy [47]; 16% antisite disorder between Fe-Al sites in $Co_2FeAl$ alloy [48]; 13% of the Ga sites are occupied by Mn atoms in $Mn_2NiGa$ alloy [6]. Therefore, we have tried to study the antisite disorders and vacancy defects with a degree of disorder (or disorder concentration or defect concentration) ranging from 3.125% to 12.50%. Here, the degree of disorder (or degree of defect) is defined as the ratio of the number of substituted (also referred to as replaced atoms) to the total number of the same kind of atoms within 64-atom supercell. When the $x$ value is chosen 0.0625 and 0.125; the disorder concentrations are 6.25% and 12.50%, respectively, for $Co_{Mn}$, $Co_{Al}$, $Mn_{Al}$, $Al_{Mn}$, $V_{Mn}$ and $V_{Al}$ disorders. For $Mn_{Co}$, $Al_{Co}$ and $V_{Co}$ disorders, the $x$ value of 0.0625 and 0.125 correspond to the disorder concentrations of 3.125% and 6.25% , respectively. Notably, for the $Mn_{Co}$, $Al_{Co}$, and $V_{Co}$ defects; disorder degree of 9.375% (corresponding to disorder at 3 Co atomic-sites out of total 32 Co atomic-sites) and 12.50% (corresponding to disorder at 4 Co atomic-sites out of total 32 Co atomic-sites) have been ignored due to limitations in computational resources. Also, for the ternary antisite disorder, the higher disorder concentrations (high values of $x$) have not been considered for the same reason.

Since describing disordered structures in terms of '$x$' values appears to be a more convenient and flexible approach, we will continue to use these specific '$x$' values to specify disorder concentration unless otherwise stated for a particular case. Prior to evaluating the physical properties of a specific disordered structure, all possible atomic configurations for that disorder concentration are considered using the supercell approach, and the full-cell relaxations were performed to identify the most stable atomic configuration (with the lowest total energy) for that specific disorder. After that, the structural, electronic, and magnetic properties of that specific disordered structure are calculated using the most stable atomic configuration. The full-cell relaxation calculations also revealed that the changes in lattice parameters and atomic positions are less than 1% for all disordered structures. Therefore, it can be concluded that the disorder leaves the cell-structure unchanged. In the following subsections, we will first discuss the formation energy, then spin polarization, and finally, the magnetic moments for the point defects.

### 3.2.1 Formation energy of the point defects

The formation energies provide information about the ease of occurrence of the disorder (*i.e.,* the order in which they can form during the experimental growth) as well as relative thermodynamic stability of the defected structures. Therefore, the relative formation energies (RFE) for the point defects *w.r.t.* $L2_1$ ordered structure, or simply referred to as the defect formation energy sometimes, have been calculated to find out the possibility of defect formation, using the following formula

$$E_{form}(X) = E_{tot}(X) - E_{tot}(L2_1) - \sum_i n_i \mu_i \quad (1)$$

where $E_{form}(X)$ and $E_{tot}(X)$ represent the relative formation energy and total energy for the defected structure (X), respectively, and $E_{tot}(L2_1)$ is the total energy for the fully ordered $Co_2MnAl$ alloy (*i.e.,* $L2_1$ structure) in the equivalent cell. Here $n_i$ is +1 (−1) for an excess (deficiency) of atom of species $i$, and $\mu_i$ is the chemical potential of the corresponding element. The $\mu_i$ for the constituting elements is taken as the ground state energy of the corresponding atoms in their bulk stable form: *hcp*-ferromagnetic Co, [001] *fcc*-antiferromagnetic Mn, and *fcc*-nonmagnetic Al. The competing crystalline phases and charge defects that might formed during defective growth are not considered. The calculated RFE using equation (1) for binary antisite disorders and vacancy defects are given in Table 2; while those for ternary antisite disorders are listed in Table 3. The negative RFE for the defected structure suggests that such defect formation is accompanied by energy release, making them likely to form spontaneously during the experimental growth. Conversely, the positive RFE for the defected structure implies that the defected structure may be formed by absorbing energy and therefore is less likely to form during growth of $Co_2MnAl$. In general, the much higher positive RFE implies that the synthesis of the defect would be less favorable in the experiments.

**Table 2:** The different disorders along with resulting structure nature; RFE, spin polarization, and minority spin band gap values for binary antisite disordered and vacancy defected structures.

| Disorder type | $x$ | Stoichiometric formula | RFE (eV) | Spin Polarization $P$ (%) | Minority spin band gap (eV) |
|---|---|---|---|---|---|
| $L2_1$ order | - | $Co_2MnAl$ | - | 76.13 | 0.66 |
| $Co_{Mn}$ antisite ($Co_{2+x}Mn_{1-x}Al$) | 0.0625 | $Co_{2.0625}Mn_{0.9375}Al$ | 0.56 | 71.56 | - |
| | 0.125 | $Co_{2.125}Mn_{0.875}Al$ | 1.13 | 67.64 | - |
| $Co_{Al}$ antisite ($Co_{2+x}MnAl_{1-x}$) | 0.0625 | $Co_{2.0625}MnAl_{0.9375}$ | 1.68 | 100.00 | 0.17 |
| | 0.125 | $Co_{2.125}MnAl_{0.875}$ | 3.38 | 100.00 | 0.19 |
| $Mn_{Co}$ antisite ($Co_{2-x}Mn_{1+x}Al$) | 0.0625 | $Co_{1.9375}Mn_{1.0625}Al$ | 0.24 | 86.73 | 0.37 |
| | 0.125 | $Co_{1.875}Mn_{1.125}Al$ | 0.48 | 88.23 | 0.37 |
| $Mn_{Al}$ antisite ($Co_2Mn_{1+x}Al_{1-x}$) | 0.0625 | $Co_2Mn_{1.0625}Al_{0.9375}$ | 1.28 | 100.00 | 0.72 |
| | 0.125 | $Co_2Mn_{1.125}Al_{0.875}$ | 2.53 | 100.00 | 0.63 |
| $Al_{Co}$ antisite ($Co_{2-x}MnAl_{1+x}$) | 0.0625 | $Co_{1.9375}MnAl_{1.0625}$ | 0.41 | 83.76 | 0.64 |
| | 0.125 | $Co_{1.875}MnAl_{1.125}$ | 0.80 | 75.53 | 0.63 |
| $Al_{Mn}$ antisite ($Co_2Mn_{1-x}Al_{1+x}$) | 0.0625 | $Co_2Mn_{0.9375}Al_{1.0625}$ | -0.67 | 71.56 | 0.55 |
| | 0.125 | $Co_2Mn_{0.875}Al_{1.125}$ | -1.41 | 87.14 | 0.24 |
| Co vacancy ($Co_{2-x}MnAl$) | 0.0625 | $Co_{1.9375}MnAl$ | 1.68 | 60.19 | - |
| | 0.125 | $Co_{1.875}MnAl$ | 3.01 | 69.08 | - |
| Mn vacancy ($Co_2Mn_{1-x}Al$) | 0.0625 | $Co_2Mn_{0.9375}Al$ | 3.04 | 54.04 | - |
| | 0.125 | $Co_2Mn_{0.875}Al$ | 5.72 | 25.60 | - |
| Al vacancy ($Co_2MnAl_{1-x}$) | 0.0625 | $Co_2MnAl_{0.9375}$ | 4.07 | 83.31 | 0.52 |
| | 0.125 | $Co_2MnAl_{0.875}$ | 7.87 | 80.72 | 0.43 |

At the thermodynamic equilibrium, the equilibrium concentration of a defect at temperature $T$ can be approximated by the equation [49,50]

$$D_{def} = N_{sites} N_{config}\, exp\left(-\frac{E_{form}}{kT}\right) \tag{2}$$

Here, $N_{sites}$ and $N_{config}$ are the number of available sites and equivalent configurations where the defect can be incorporated. However, using equation (2), only a qualitative estimate of the defect concentration is possible, as equation (2) is strictly valid in the limit of very low concentrations and greatly exceeds when considering more than a few percent.

*For the binary antisite disorders and vacancy defects with x = 0.0625*; RFE is positive except for $Al_{Mn}$ antisite disorder. The $Al_{Mn}$ antisite disorder has negative RFE; therefore, $Al_{Mn}$ disorder is most likely to be formed during the experimental growth of $Co_2MnAl$, and it is supposed to be formed spontaneously, as suggested by equation (2). For rest, the decreasing order for probability of formation is: $Mn_{Co}$ , $Al_{Co}$ , $Co_{Mn}$, $Mn_{Al}$, $Co_{Al}$, $V_{Co}$, $V_{Mn}$ and $V_{Al}$. Although, the $Co_{Al}$ antisite and $V_{Co}$, $V_{Mn}$, and $V_{Al}$ vacancies defected structure have very high positive RFE , their occurrence cannot be excluded. They are expected to survive with a small density according to equation (2). *While increasing the disorder concentration, i.e., with x = 0.125*; the RFE nearly doubles for all cases. Such changes in RFE are naturally established here due to the almost localized and isolated nature of disorders, which will be discussed later in Sections 3.2.2 & 3.2.3. These results for RFE exhibit a similar trend for the cases that have been studied in Ref. 28 for $Co_2MnAl$. They are also consistent with the other first-principal findings for $Co_2MnSi$ alloy [51], where the Si is analogous to Al in $Co_2MnAl$ alloy.

**Table 3:** The different ternary antisite disorders along with the resulting structure nature; corresponding stoichiometric formula, RFE, spin polarization, and minority spin band gap values for the ternary antisite disordered structures.

| **Disorder type** | **Structure nature** | ***x*** | **Stoichiometric formula** | **RFE (eV)** | **Spin Polarization *P* (%)** | **Minority spin band gap (eV)** |
|---|---|---|---|---|---|---|
| $Mn_{Co} + Al_{Co}$ ($Co_{2-2x}Mn_{1+x}Al_{1+x}$) | Co-deficient | 0.0625 | $Co_{1.875}Mn_{1.0625}Al_{1.0625}$ | 0.60 | 84.50 | 0.52 |
| $Co_{Mn} + Al_{Mn}$ ($Co_{2+x}Mn_{1-2x}Al_{1+x}$) | Mn-deficient | 0.0625 | $Co_{2.0625}Mn_{0.875}Al_{1.0625}$ | -0.10 | 66.99 | - |
| $Co_{Al} + Mn_{Al}$ ($Co_{2+x}Mn_{1+x}Al_{1-2x}$) | Al-deficient | 0.0625 | $Co_{2.0625}Mn_{1.0625}Al_{0.875}$ | 2.75 | 100.00 | |

For the case of *ternary antisite disorders*; the Mn-deficient structure resulting from ($Co_{Mn}$+$Al_{Mn}$)-antisite disorder exhibits the negative RFE, therefore they are likely to be formed spontaneously during experimental growth of $Co_2MnAl$; similar to $Al_{Mn}$ antisite disorder. Meanwhile, the Co-deficient (resulting from ($Mn_{Co}$+$Al_{Co}$)-antisite disorder) and Al-deficient (resulting from ($Co_{Al}$+$Mn_{Al}$)-antisite disorder) structures have moderate and very high RFE, respectively. Consequently, the likelihood of the occurrence is significantly reduced for Co-deficient structure, and almost negligible for Al-deficient structure. As noted from Tables 2 & 3, the RFE of the ternary antisite disorder can be approximated by summing the RFE of individual binary antisite involved, which is the direct signature of isolation of binary antisite disorder (significant distance between the substituted atoms) inside the energetically favored ternary antisite disorder structure within the 64-atom supercell.

When considering the formation of various disorders, it is important to note that the RFE values listed in Tables 2 & 3 are computed by considering the constituent atoms in their bulk stable form (under host-rich conditions), and at absolute zero temperature and absolute zero pressure. However, a similar qualitative behavior (order of formation) should be observed under realistic experimental thermodynamic conditions (*i.e.*, at non-zero temperature and pressure). It is also crucial to stress about the negative RFE for $Al_{Mn}$ binary antisite disordered structures ($Co_2Mn_{0.9375}Al_{1.0625}$ and $Co_2Mn_{0.875}Al_{1.125}$) and for ($Co_{Mn}$+$Al_{Mn}$) ternary antisite disordered structure ($Co_{2.0625}Mn_{0.875}Al_{1.0625}$), which signify that these disordered structures are relatively more likely to be formed compared to the pure crystalline system during the growth of $Co_2MnAl$, as also supported by equation (2). However, there still exists a lack of control over the thermodynamics for the growth process, and also of the well-defined experimental method to probe such defects. For such cases, the formation of the mixed phase due to antisite disorder has been reported for many Heusler alloys [52,53]. To resolve this question unambiguously, advanced first-principles approaches are necessary; like in Ref. 44 or the ab initio atomistic thermodynamic calculations, which are beyond the scope of the present study [44].

### 3.2.2 Effect of point defects on the electronic properties of $Co_2MnAl$

Next, we pay attention to the effects of point defects (*i.e.,* binary antisite disorders, vacancy defects, and ternary antisite disorders) on the electronic properties of $Co_2MnAl$ alloy. Before describing the influence of the point defects on the electronic properties, let us first discuss some important features of the density of states (DOS) of ideal (or $L2_1$ ordered) $Co_2MnAl$ alloy. Like the DOS plot for other Co-based full Heusler alloys, DOS plot of $Co_2MnAl$ exhibits the peak and valley characteristics resulting from the *d*-orbitals localization and van Hue singularities, as seen in Figure S1(a) of the supplementary material [54]. The $E_F$ lies in the spin valley for the majority spin channel (or the spin-up DOS) and in the pseudogap for the minority spin channel (or the spin-down DOS). As a result, $Co_2MnAl$ is not a perfect half-metal; instead, it has a very high spin polarization of ~76.13% with nearly half-metallic characteristics. $Co_2MnAl$ has a wide minority spin energy gap or the Slater Pauling (SP) valley of 0.66 eV, located just above $E_F$. The origin of this minority gap has already been discussed in detail in literature and is understood to be associated with the covalent hybridization between transition metal elements.

To study the impact of the disorder on the electronic properties, the total DOS of the disordered structures are compared to the ideal case. In the presence of point defects, the majority channel DOS remains nearly the same as the *L2₁*-ordered structure and the changes happen in the minority spin channel only, near the $E_F$, as depicted in Figures 1 & 2. The extent of these changes depends on the specific type of disorder. For most of the considered disordered structures, the SP valley in the minority is nearly unscathed, with only minor changes. However, in some cases, substantial changes have also occurred in the minority DOS, leading to the emergence of new states in the minority spin channel. In addition to this, the minority gap width ($E_c$-$E_v$) also changes significantly for all disorders (see Tables 2 & 3); while the Fermi energy (equivalently change in the minority gap center {($E_c$+$E_v$)/2} or shifting in SP valley) experiences minor but important modifications (Tables S1 & S2 in the supplementary material). Here, $E_c$ and $E_v$ represent the minority valence band edge and minority conduction band edge near $E_F$. The above-discussed alterations in the DOS plots can be attributed to a multitude of factors: the reason that the changes happened only in the minority spin channel of the ideal DOS, after incorporating the disorder, can be explained as a step to minimize the band energy of the disordered structure, as discussed by Faleev *et al.* for similar Heusler alloys [54]. Further, the changes in the minority gap width, Fermi energy, and the emergence of new states in the minority spin channel emerge from changes in the hybridization between the constituting atoms of the disordered structures.

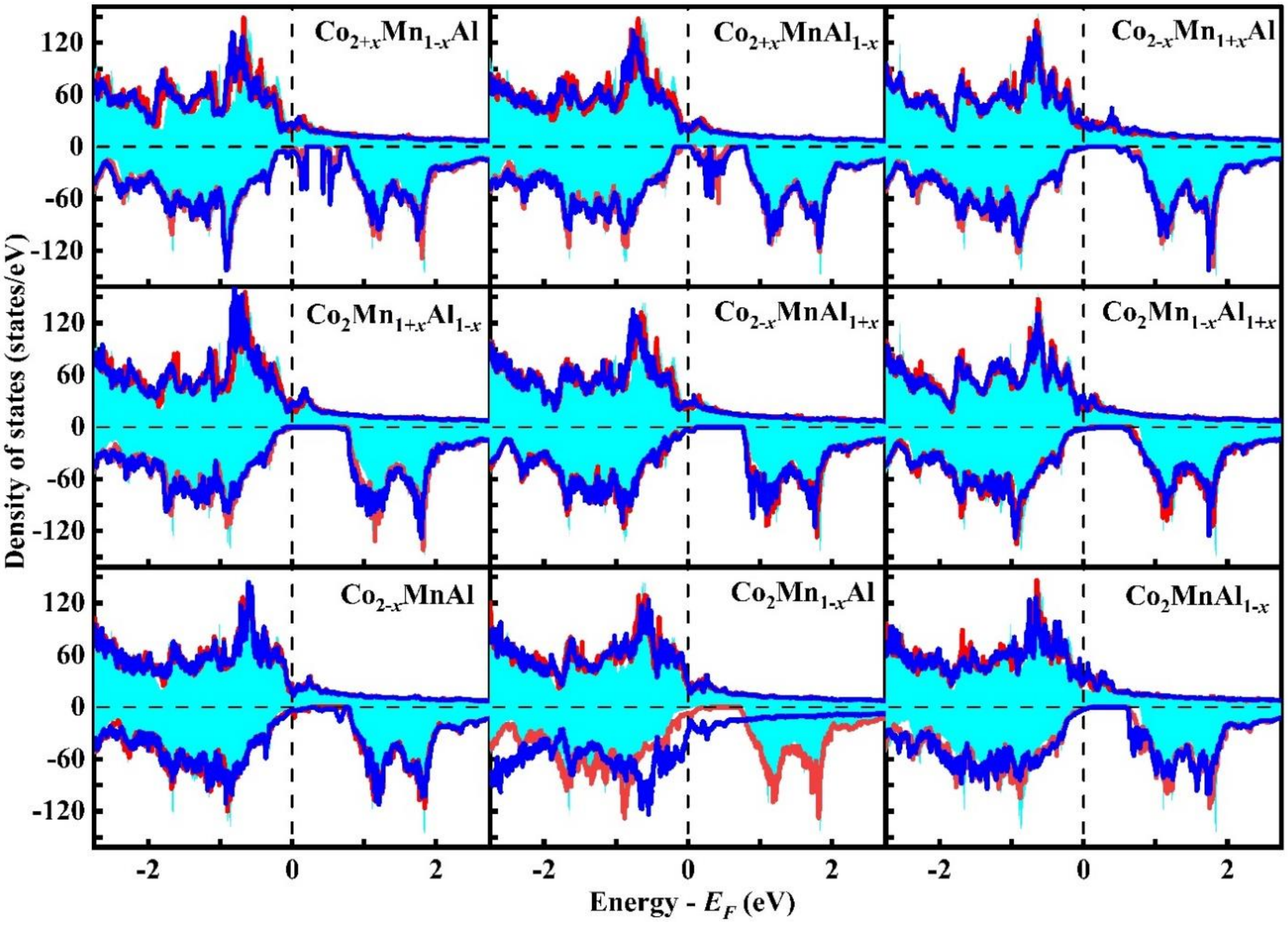

**Figure 1:** The DOS plots of various binary antisite disordered and vacancy defected structures. The shaded cyan area represents the DOS of ideal $Co_2MnAl$, while the solid red and blue lines represent DOS for $x$ = 0.0625 and $x$ = 0.125, respectively. For the disordered structure, the stochiometric formula is given in the insight of each graph.

After that, the spin polarization for the disordered structures is determined using the formula, $P = (D_1 - D_2)/(D_1+D_2)$, where $D_1$ and $D_2$ represent the spin-up and spin-down DOS at the $E_F$, respectively. The alternations in DOS, as discussed above, lead to changes in $P$ for the disordered structure. The calculated $P$ and minority spin-gap (real gap for structures having 100% spin polarization and pseudogap for others) for the binary antisite disordered structures, vacancy defected structures, and ternary antisite disordered structures are listed in Tables 2 & 3. The antisite disordered (binary as well ternary) structures exhibit relatively high spin polarization (compared to the ordered structure). On the other hand, for the vacancy defected structures (except the Al-vacancies), a significant reduction in $P$ is observed.

*For binary antisite disorders and vacancy defects with the disorder concertation of* $x = 0.0625$; the minor modifications in DOS are observed for the Mn-rich structures ($Mn_{Co}$ and $Mn_{Al}$ antisite), Al-rich structures ($Al_{Co}$ and $Al_{Mn}$ antisite) and Al-vacancy; which can be described in terms of changes in the minority gap width and the shifting of the SP valley towards the high binding energies (Figure 1). As a result, due to the presence of the smaller number of minority DOS at $E_F$ as compared to the ideal case, an increment in $P$ is observed for the above-mentioned disordered structures. Remarkably, for the $Mn_{Al}$ disordered structure, the absence of the minority spin states at $E_F$ results in half-metallic nature. Additionally, the $Mn_{Al}$ disordered structure possesses a wide minority energy gap of 0.72 eV, even larger than the $L2_1$-ordered structure; making the $Mn_{Al}$ disordered structure more useful than the $L2_1$-ordered structure for spintronic applications. Meanwhile, in the case of Co-rich structures ($Co_{Al}$ and $Co_{Mn}$ antisite), $V_{Co}$, and $V_{Mn}$; significant alternations in DOS are observed along with the new bands arise within the minority gap. For the $Co_{Al}$ antisite disorder, this modification is such that the $E_F$ falls in the energy gap in minority spin, resulting in 100% spin polarization for $Co_{Al}$ disordered structure as well; whereas for $Co_{Mn}$ antisite disorder, the complete closing of the minority spin gap takes place from the emergence of new bands at $E_F$, therefore a slight reduction (~71.56%) in $P$ is observed for $Co_{Mn}$ disordered structure. For the Co- and Mn-monovacancy defects, the minority DOS are substantially affected, leading to the formation of a continuous band. This band is more intense in the presence of the Mn-monovacancy than the Co-monovacancy, resulting in a large reduction in spin polarization (~54.04 %) for the Mn-monovacancy defected structure.

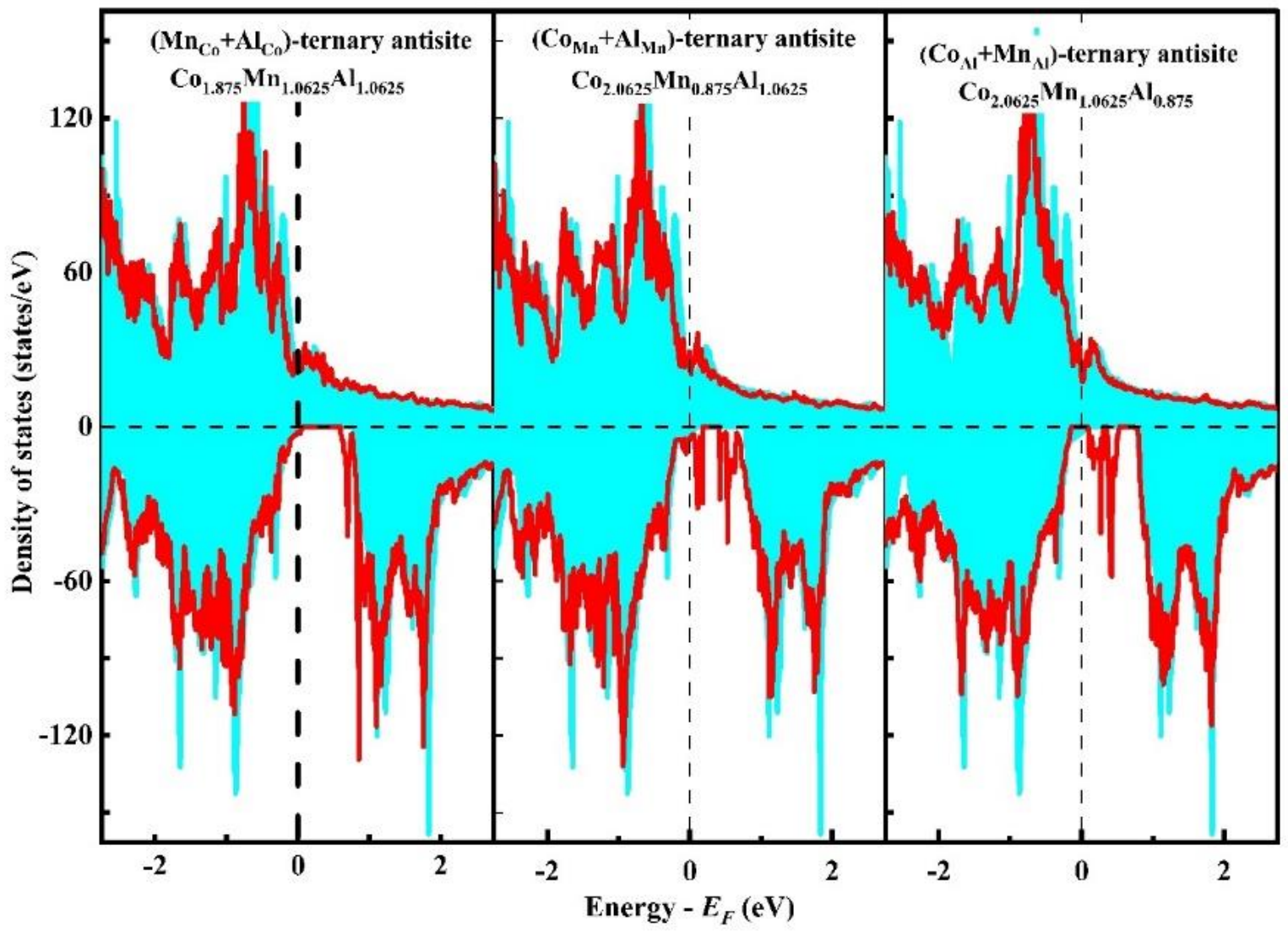

**Figure 2:** The DOS plots of various ternary antisite disordered structures. The shaded cyan area represents the DOS of ideal $Co_2MnAl$, while the solid red for $x = 0.0625$. For the disordered structure, the stochiometric formula is given insight of each graph.

*Upon increasing the disorder concentration (i.e., for x = 0.125)*; $Mn_{Al}$ and $Co_{Al}$ antisite disordered structures retain their half-metallic nature. On the other hand, for the $Mn_{Co}$ and $Al_{Mn}$ antisite disordered structures, $P$ increased; whereas for the $Co_{Mn}$ and $Al_{Co}$ antisite disordered structure, $P$ is reduced. With the change in the disorder concentrations, the minority band gap values for the $Mn_{Al}$ and $Co_{Al}$ disordered structures changed to 0.63 eV and 0.19 eV, respectively, along with 100% $P$, making them beneficial in the presence of disorder, specially in the case of $Mn_{Al}$ disorder. *For the case of di-vacancies (i.e., vacancy defects with x = 0.125)*, the impact is much more intense as compared to the mono-vacancies. For Mn di-vacancy, the complete shrinking of the minority pseduogap takes place (see Figure 1), leading to a large reduction in spin polarization (~25%). However, for the Co and Al di-vacancies, the effect on DOS and $P$ is very alike in their mono-vacancies. Therefore, for Co and Al di-vacancies, spin polarization is nearly the same as that of mono-vacancies.

*Next, for the ternary antisite disorders*, DOS plots are given in Figure 2. For the Co-deficient structures, there are marginal changes in the DOS shape. In contrast, for the Mn- and Al-deficient structures, the emergence of new states in the minority gap is accompanied by changes in the gap width and shifting of minority DOS (Figure 2). In the Co-deficient structure (or ($Mn_{Co}$+$Al_{Co}$)-ternary disordered structure), reduced minority DOS at $E_F$ leads to the enhancement in $P$. For the Al-deficient structure (or ($Co_{Al}$+$Mn_{Al}$)-ternary disordered structure); the change in DOS is in such a way that $E_F$ falls in real gap like in the $Co_{Al}$ antisite disordered structures, resulting in 100% $P$; whereas in the Mn-deficient structure (or the ($Co_{Mn}$+$Al_{Mn}$)-ternary antisite disordered structure), the minority spin states at $E_F$ get intensified compared to the ideal structure, leading to the reduction in $P$.

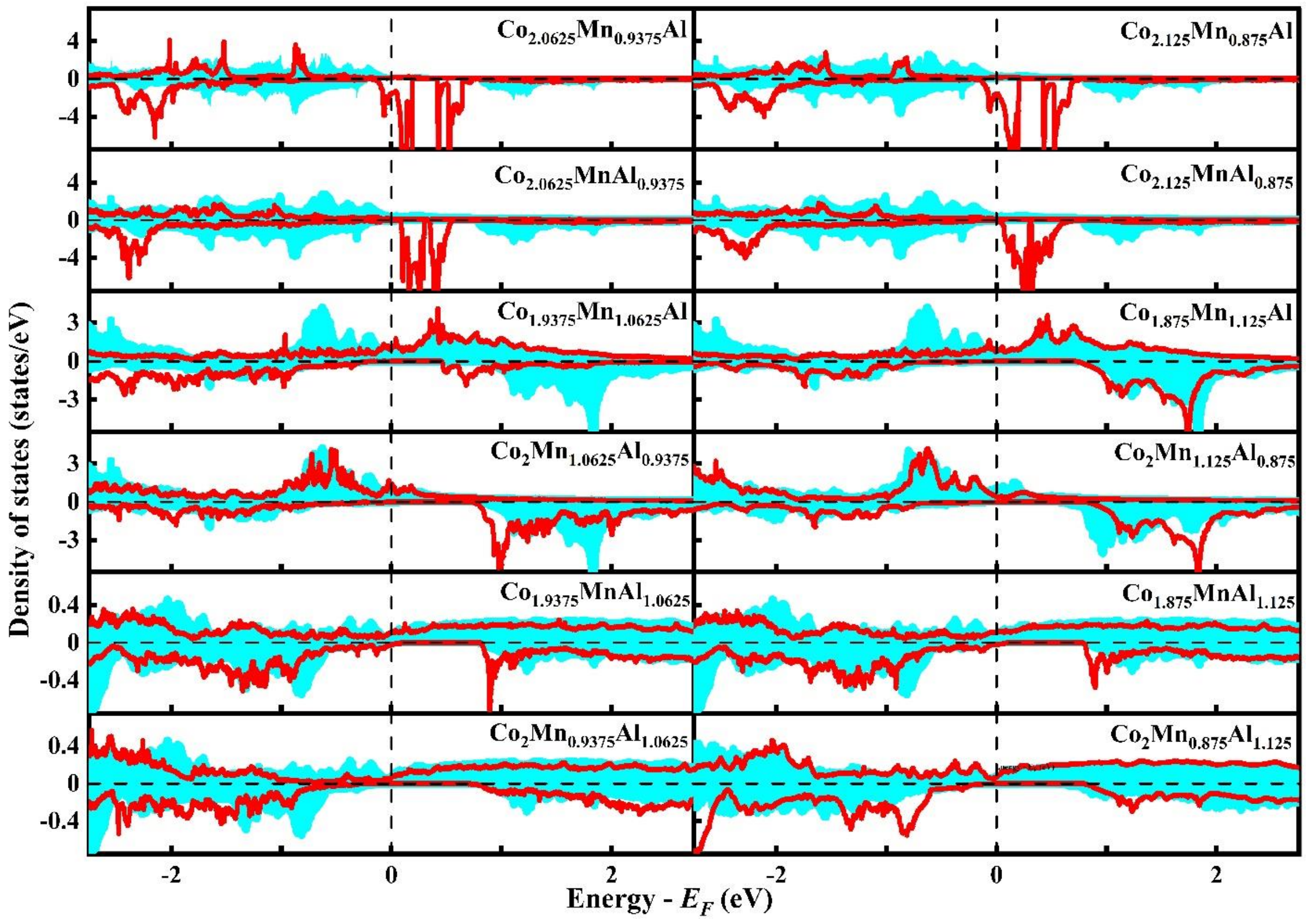

**Figure 3:** The PDOS plots for the disordered (substituted) atom in binary antisite disorders. The solid red lines represent PDOS for the disorder atom, the shaded cyan area represents the PDOS for the corresponding atom in the ideal $Co_2MnAl$ alloy. The corresponding stoichiometric formula for the disordered structure is given in the inset of each graph. It was found that where two atoms were substituted, both substituted atoms have similar PDOS in the case of the two atoms are substituted; therefore, PDOS for only one atom is shown here. (Since, in the case of vacancy defects, the atom is absent at the defect site; therefore, PDOS plots for vacancy defects are not provided.)

For further exploring the origin of observed changes in the DOS (and hence in spin polarization) of disordered structures, particularly the emergence of new states in SP valley; the partial density of states (PDOS) for the substituted (or disordered or replaced) atoms and its surrounding atoms have been calculated. PDOS plot for disordered atoms in Figures 3 (binary antisites) & 4 (ternary antisites) reveal that the major changes in DOS near $E_F$ arise from the disordered atom in the disordered structures; whereas the contributions from the rest of the atoms within unit-cell are small and can be neglected (PDOS for other atoms not shown). This change between PDOS of the disordered atom in the disordered structure from PDOS at the perfect site in $Co_2MnAl$ is due to the different symmetry at which they sit in case of disordered structure. Thus, it can be summarized that the effect of binary antisite and ternary antisite disorder on the electronic properties are localized in nature and the DOS for and binary antisite disordered as well for ternary antisite disordered structure can be almost expressed as a superposition of the PDOS of the disordered atom(s) and the ideal atoms PDOS. On the other hand, PDOS plots for the vacancy-defected structures similar to Figures 3 and 4, are not possible. This is due to the absence of atom(s) at defect site(s). However, for their analysis, PDOS plots from the other atoms in the defected structures have been utilized (not shown for brevity). In the vacancy defected structures, the changes in PDOS were found to survive for several neighbors of vacancy site(s), almost spread throughout the unit-cell. This suggests that the vacancy defects have long-range effects on the electronics properties and lead to the changes in hybridization up to next to the nearest neighbors. This also makes the electronic properties of disordered alloy vulnerable to the variations in vacancy defect concentrations.

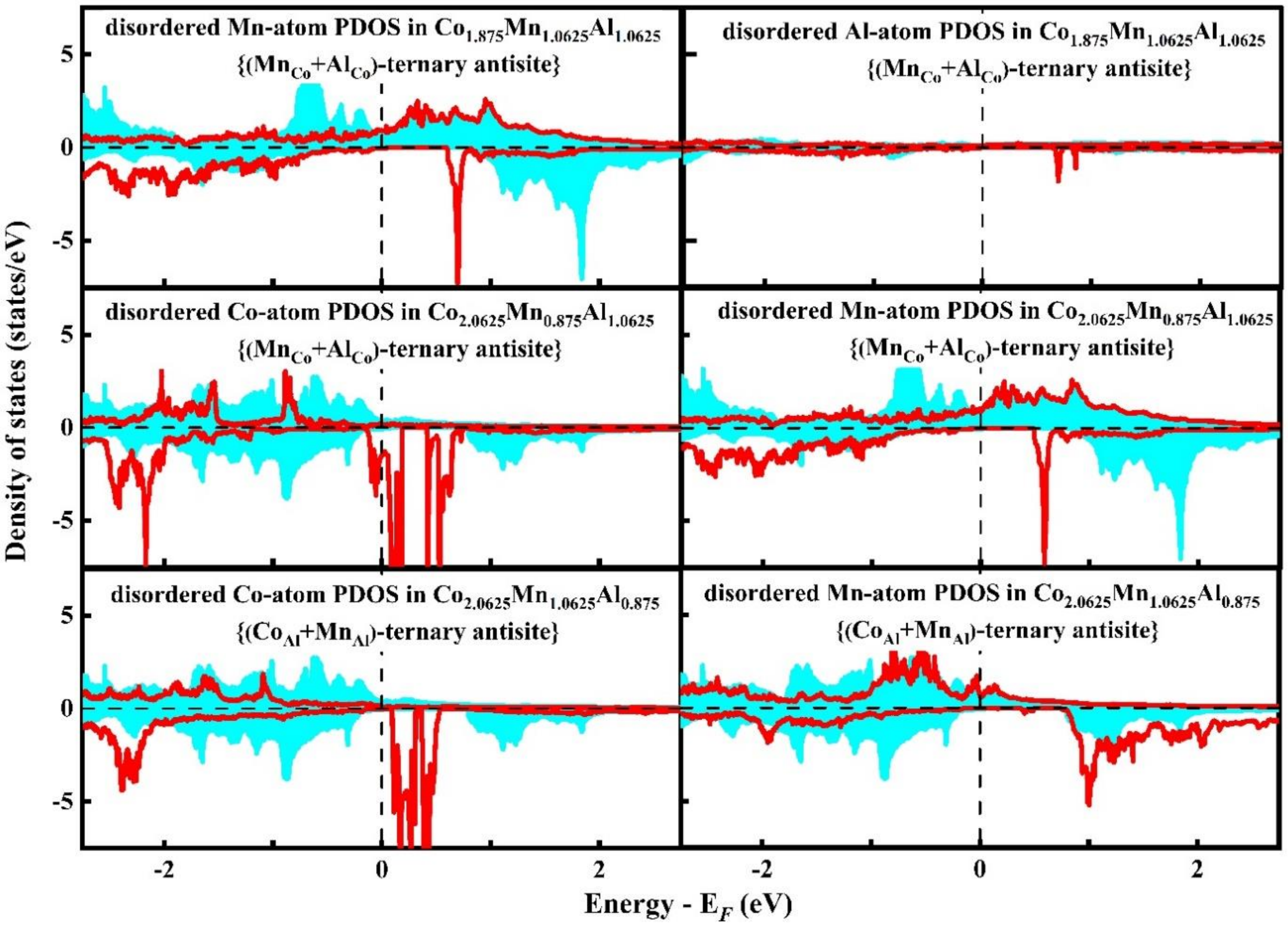

**Figure 4:** The PDOS plots for the disordered (substituted) atom in ternary antisite disorders. The solid red lines represent PDOS for disorder atom, the shaded cyan area represents the PDOS for the corresponding atom in ideal $Co_2MnAl$ alloy.

Finally, let us compare these findings with the available results in Refs. 27 & 28. As stated in the introduction section; Ref. 28 discussed the pseudogap, minority states across $E_F$ and $P$ for the $Co_{Mn}$, $Mn_{Co}$, $Mn_{Al}$ and $Al_{Mn}$ antisite disordered structures at different disorder concentrations (quantified). In Ref. 27, the authors reported the effect of $Al_{Mn}$, $Mn_{Al}$ and $Mn_{Co}$ antisite disorder, and vacancy defect on the pseudogap and minority states across $E_F$; however, they did not provide the spin polarization values. Our results for the changes in $P$ and minority states across $E_F$ with disorder concentrations for binary antisite disordered structures closely resemble the trends seen in the limited data from Refs. 27 & 28. The slight disparities in the results can be attributed to variations in the supercell size and differences in the stoichiometric formula. However, in the case of vacancy defects, our results concerning the alterations in minority DOS differ from the observations in Ref. 27, which reveal that Co vacancies strongly affect the minority states, whereas the Mn vacancies have a moderate effect. Conversely, present findings indicate that Mn vacancies have the most substantial effect, while Co vacancies yield a moderate influence. This disparity in the results may be ascribed to several factors, such as the non-localized nature of the impact of vacancy defects, slight variations in the stoichiometric formula, and differences in the calculation methods. The findings reported in Ref. 27 were computed using the full-potential local-orbital method with LDA for exchange, along with Coherent Potential Approximation method to simulate disorder randomly.

### 3.2.3 Effect of point defects on magnetic properties of $Co_2MnAl$

After that, we will discuss the magnetic moment of disordered structures. Recall that $Co_2MnAl$ in the bulk phase ($L2_1$-ordered structure) possesses a total magnetic moment of 4.02 $\mu_B$/*f.u.* equivalently, a magnetic moment of 64.32 $\mu_B$ in the 64-atoms supercell. As the defected structures are modeled using 64-atom supercells, which are also the smallest unit cell for point defects, the total moment for the $L2_1$ ordered structure and various disordered structures is described using the 64-atom cell. For the disordered structures; the stoichiometric formula, total magnetic moment, atomic moment of the disordered atom(s), and estimated total magnetic moment as per Slater Pauling rule [55] are gathered in Tables 4 & 5. The total magnetic moment as Slater Pauling rule is estimated using the formula (Z-16×24), where Z as the total valence electrons in 64-atom supercell. The compositional formula for the disordered structures is also provided for reader's reference. It is important to note that the Slater Pauling rule is most accurate for perfect half-metallic HAs. For nearly half-metallic HAs, it provides a reasonably close estimate.

The total magnetic moment of the disordered structures exhibits slight deviations from the total moment of ideal structure, as illustrated in Tables 4 & 5. The extent of these deviations depends upon the specific disorder present in sample, like in the case of $P$. The total magnetic moment for an electronic structure is given by $\mu_B$×(number of spin-up electrons – number of spin-down electrons). Since, in the context of disordered structures, the substitution of atoms induces a change in the number of total valence electrons;

therefore, a change in the total magnetic moment is also expected. Also, as the valence electron count changes due to the substituted atom(s), the atomic magnetic moment is expected to be changed only for these specific atom(s) from the corresponding ideal atoms' moments. However, it is found that the neighboring atoms to the substituted atoms also experience variations in their magnetic moments, and this phenomenon is specific to the type of disorder present. To gain deeper insight into the magnetic moments for other atoms, the local magnetic moment for all atoms within the 64-atom supercell has been mapped as the distance from disorder atom(s) in Figures S2 (a)-(r) & S3(a)-(c). These Figures clearly represent the variations in atomic magnetic moments within the unit cell. These changes in the atomic magnetic moments can be attributed to the changes in exchange splitting resulting from changes in hybridization. In the following, the magnetic moment for the different point defects, each with different concentration, are presented.

**Table 4:** The different binary antisite and vacancy disordered structures and their total magnetic moments (calculated and as per the SP rule), and magnetic moments for disordered atom(s). $X_d$ denotes the magnetic moment for the substituted atom(s).

| Disorder structure | $x$ | Composition formula | Total magnetic moment ($\mu_B$/cell) | | Magnetic moment for substituted atom(s) $X_d$ (in $\mu_B$) |
|---|---|---|---|---|---|
| | | | Calculated | SP Rule | |
| *L2₁* ordered structure | | $Co_{32}Mn_{16}Al_{16}$ ($Co_2MnAl$) | 64.32 | 64 | - |
| $Co_{Mn}$ antisite disorder ($Co_{2+x}Mn_{1-x}Al$) | 0.0625 | $Co_{33}Mn_{15}Al_{16}$ ($Co_{2.0625}Mn_{0.9375}Al$) | 65.46 | 66 | Co: 1.93 |
| | 0.125 | $Co_{34}Mn_{14}Al_{16}$ ($Co_{2.125}Mn_{0.875}Al$) | 66.74 | 68 | Co: 1.93<br>Co: 1.93 |
| $Co_{Al}$ antisite disorder ($Co_{2+x}MnAl_{1-x}$) | 0.0625 | $Co_{33}Mn_{16}Al_{15}$ ($Co_{2.0625}MnAl_{0.9375}$) | 70.00 | 70 | Co: 2.07 |
| | 0.125 | $Co_{34}Mn_{16}Al_{14}$ ($Co_{2.125}MnAl_{0.875}$) | 76.00 | 76 | Co: 2.08<br>Co: 2.08 |
| $Mn_{Co}$ antisite disorder ($Co_{2-x}Mn_{1+x}Al$) | 0.0625 | $Co_{31}Mn_{17}Al_{16}$ ($Co_{1.9375}Mn_{1.0625}Al$) | 62.11 | 62 | Mn : -2.09 |
| | 0.125 | $Co_{30}Mn_{18}Al_{16}$ ($Co_{1.875}Mn_{1.125}Al$) | 60.08 | 60 | Mn : -2.09<br>Mn : -2.09 |
| $Mn_{Al}$ antisite disorder ($Co_2Mn_{1+x}Al_{1-x}$) | 0.0625 | $Co_{32}Mn_{17}Al_{15}$ ($Co_2Mn_{1.0625}Al_{0.9375}$) | 68.00 | 68 | Mn: 2.45 |
| | 0.125 | $Co_{32}Mn_{18}Al_{14}$ ($Co_2Mn_{1.125}Al_{0.875}$) | 72.00 | 72 | Mn : 2.42<br>Mn: 2.42 |
| $Al_{Co}$ antisite disorder ($Co_{2-x}MnAl_{1+x}$) | 0.0625 | $Co_{31}Mn_{16}Al_{17}$ ($Co_{1.9375}MnAl_{1.0625}$) | 64.44 | 58 | Al: -0.05 |
| | 0.125 | $Co_{30}Mn_{16}Al_{18}$ ($Co_{1.875}MnAl_{1.125}$) | 64.89 | 52 | Al: -0.05<br>Al: -0.05 |
| $Al_{Mn}$ antisite disorder ($Co_2Mn_{1-x}Al_{1+x}$) | 0.0625 | $Co_{32}Mn_{15}Al_{17}$ ($Co_2Mn_{0.9375}Al_{1.0625}$) | 60.24 | 60 | Al: -0.05 |
| | 0.125 | $Co_{32}Mn_{14}Al_{18}$ ($Co_2Mn_{0.875}Al_{1.125}$) | 56.71 | 56 | Al: -0.04<br>Al: -0.04 |
| Co vacancy ($Co_{2-x}MnAl$) | 0.0625 | $Co_{31}Mn_{16}Al_{16}$ ($Co_{1.9375}MnAl$) | 63.15 | 55 | - |
| | 0.125 | $Co_{30}Mn_{16}Al_{16}$ ($Co_{1.875}MnAl$) | 63.32 | 46 | - |
| Mn vacancy ($Co_2Mn_{1-x}Al$) | 0.0625 | $Co_{32}Mn_{15}Al_{16}$ ($Co_2Mn_{0.9375}Al$) | 59.33 | 57 | - |
| | 0.125 | $Co_{32}Mn_{14}Al_{16}$ ($Co_2Mn_{0.875}Al$) | 55.99 | 50 | - |
| Al vacancy ($Co_2MnAl_{1-x}$) | 0.0625 | $Co_{32}Mn_{16}Al_{15}$ ($Co_2MnAl_{0.9375}$) | 61.26 | 61 | - |
| | 0.125 | $Co_{32}Mn_{16}Al_{14}$ ($Co_2MnAl_{0.875}$) | 58.35 | 58 | - |

*For the binary antisite disorders with $x = 0.0625$*; Al-poor structures resulting from binary antisite disorder ($Co_{Al}$ and $Mn_{Al}$ antisite disorders) show the increased total magnetic moment; whereas Al-rich structure (resulting from $Al_{Mn}$ antisite disorder) shows decreased magnetic moment. The rest of the disordered structures; *viz.,* containing $Mn_{Co}$, $Al_{Co}$, and $Co_{Mn}$ antisite disorder, have mild effects on the total magnetic moment, and the total magnetic moment remains nearly unchanged for them. The substantial changes in the total magnetic moment for disordered structures containing $Co_{Al}$, $Mn_{Al}$, and $Al_{Mn}$ antisite disorders are due to the significant difference in valence electron count for the dopant and host atoms. Notably, Co, Mn and Al have 9, 7 and 3 valence electrons, respectively (according to the pseudopotential configurations used for calculations). Based upon the valence electrons count, a similar behavior should also be observed for the $Al_{Co}$ disordered structure *i.e.*, decreased total magnetic moment compared to the ideal structure. However, the small increments in several neighboring atoms' atomic magnetic moments (AMMs) accumulate and thus yield a total magnetic moment close to that of the ideal structure, as seen in Figures S2(j)-(k). Furthermore, it can also be observed from Figures S2 that for all considered binary antisite disorders, the AMMs at regular sites are barely changed or undergo negligible small change. This observation indicates that the influence of antisite disorder on the magnetic properties is localized in nature.

**Table 5:** Various types of ternary antisite disordered structures and their total magnetic moments (calculated and as per the SP rule), and magnetic moments for disordered atom(s). $X_d$ denotes the magnetic moment for the substituted atom(s).

| **Disorder Type** | ***x*** | **Composition formula** | **Total magnetic moment ($\mu_B$/cell)** | | **Magnetic moment for substituted atom(s) $X_d$ (in $\mu_B$)** |
|---|---|---|---|---|---|
| | | | **Calculated** | **SP rule** | |
| $Mn_{Co}$ + $Al_{Co}$ ($Co_{2-2x}Mn_{1+x}Al_{1+x}$) | 0.0625 | $Co_{30}Mn_{17}Al_{17}$ ($Co_{1.875}Mn_{1.0625}Al_{1.0625}$) | 62.19 | 56 | Mn: -2.12<br>Al: -0.03 |
| $Co_{Mn}$ + $Al_{Mn}$ ($Co_{2+x}Mn_{1-2x}Al_{1+x}$) | 0.0625 | $Co_{33}Mn_{14}Al_{17}$ ($Co_{2.0625}Mn_{0.875}Al_{1.0625}$) | 61.56 | 62 | Co : 1.93<br>Al : -0.05 |
| $Co_{Al}$ + $Mn_{Al}$ ($Co_{2+x}Mn_{1+x}Al_{1-2x}$) | 0.0625 | $Co_{33}Mn_{17}Al_{14}$ ($Co_{2.0625}Mn_{1.0625}Al_{0.875}$) | 74.00 | 74 | Co : 2.07<br>Mn : 2.49 |

*For the binary antisite disordered structures corresponds to x = 0.125*, owing to the considerable far distance between substituted atoms in the disordered structure and the localized nature of binary antisite disorder with $x = 0.0625$ (for the magnetic moments), the magnetic moment for the disordered structure changes in a similar manner as in the case of binary antisite disordered structures with $x = 0.0625$; and changes are amplified in the proportion of the disorder concentration (*i.e.,* doubled). Also, the effect of disorder on magnetic moments remained localized, as illustrated in Figures S2(a)-(l). The magnetic moment values for $Co_{Mn}$, $Mn_{Al}$, and $Al_{Mn}$ antisite disordered structures are closely aligned and exhibit a similar trend with a change in disorder concentration (*i.e.,* change in a similar way), with the results reported in Refs. 27 and 28, wherever possible. This alignment validates our findings and can be attributed to the localized nature of antisite defects, which lead to comparable outcomes despite having nearly identical stoichiometric formulas. Further, any minute differences observed can be attributed to slight variations in the stoichiometric formula considered in Refs. 27 and 28.

*For the vacancy defects with x = 0.0625, corresponding to the mono-vacancies*, the absence of an atom within the lattice leads to a notable decrease in the total magnetic moment. This decrement is evident, as the absence of an atom leads to a decrease in the total valence count. Furthermore, as illustrated in Figures S2(m)-(r), unlike the case of the binary antisite, the AMM map for the vacancy defected structures is complicated in nature, and neighboring atoms also make significant contributions to change in total magnetic moment. As a result, the calculated magnetic moment values for the vacancy defected structures differ significantly from the SP rule. The most substantial changes occur in the presence of $V_{Mn}$ and $V_{Al}$, where the magnetic moment decreases significantly. In contrast, the presence of $V_{Co}$ has marginal effect on the total magnetic moment, leaving it nearly the same as the ideal structure. For the vacancy defects, considerable fluctuations in the magnetic moments of distant neighboring atoms indicate that the vacancy defects have non-localized effect on the AMMs. Because of this non-localized nature (or long-range survival) of the vacancy defects impact, the change in the magnetic moments happens arbitrarily on increasing vacancy defects concentration (*i.e.*, *with x = 0.125 or for the di-vacancies*). As observed in Figures S2(m)-(r), the alterations in AMMs extend to $6^{th}$ nearest neighbors for Co mono-vacancy, up to the $3^{rd}$ nearest neighbors for Mn mono-vacancy and up to the $2^{nd}$ nearest neighbors for Al mono-vacancy. In the scenario of di-vacancies (*i.e.,* with $x$ = 0.125), these effects reach out to the $4^{th}$, $4^{th}$, and $2^{nd}$ nearest neighbors in the case of Co, Mn, and Al atoms' di-vacancies, respectively.

*Finally, in the case of the ternary antisite disorders*; the total magnetic moment for the Co-deficient structure (($Mn_{Co}$+$Al_{Co}$)-ternary antisite) and Mn-deficient structure (($Co_{Mn}$+$Al_{Mn}$)-ternary antisite) is nearly identical to the ideal structure's magnetic moment, due to almost same valence electrons; whereas for the Al-deficient structure (resulting from ($Co_{Al}$+$Mn_{Al}$)-ternary antisite) exhibits an increased magnetic moment due to the significant alternations in valence electrons count. For all considered ternary antisite disorders, the impact of the disorder on magnetic properties is also localized, similar to binary antisite disorders, as demonstrated in Figures S3(a)-(c) of supplementary material.

Hence, within the context of point defects, it can be stated that both binary and ternary antisite disorders result in localized alterations in magnetic moments; whereas the vacancy defects lead to complex change in magnetic moments, which is very sensible to their concentrations. This behavior closely resembles that observed for electronic properties (spin polarization) and is also expected since both are linked to the PDOS. Thus, based on the magnetic moment and DOS analysis (Section 3.2.2), it can be inferred that point defects act as a minor perturbation in the case of antisite disorder, whether binary or ternary, but it is notable when it comes to vacancy defects.

Another noteworthy observation in the overall picture is that the SP rule is followed by the disordered structures having 100% polarization *viz.,* Al-deficient structure from binary and ternary antisite disorder structures ($Co_{Al}$ and $Mn_{Al}$ binary antisite structures, and ($Co_{Al}$+$Mn_{Al}$)-ternary antisite structure; *c.f.* Tables 4 & 5). This reaffirms the half-metallicity for these disordered structures, as discussed in Section 3.2.2. Additionally, for the remaining disordered structures, the total magnetic moment closely corresponds to the predictions of the Slater-Pauling rule, except in the case of $Al_{Co}$ antisite and vacancy defects. For $Al_{Co}$ disorder, some small positive contributions from neighboring atoms in the disordered structure yield a total magnetic moment that closely resembles that of the ideal structure, as discussed earlier. In vacancy defects, deviations from the SP rule can be ascribed to the alterations in AMMs throughout the unit cell, originating from the non-localized nature of vacancy defect impacts.

Notably, the most interesting changes were observed in the Co-deficient structures originating from $Mn_{Co}$ binary antisite disorder and ($Mn_{Co}$+$Al_{Co}$)-ternary antisite disorder, for all considered $x$ values. The presence of $Mn_{Co}$ antisite among these structures results in the antiferromagnetic coupling between the antisite-Mn atom and neighboring ideal atoms (Co and Mn). Since these disordered structures also possess very high spin polarization (as shown in Tables 2 & 3), it can be concluded that the creation of $Mn_{Co}$ antisite leads to half-metallic anti-ferrimagnetism (HMFi), which is highly desirable for spintronics applications. This beautiful phenomenon has also been theoretically observed for many other Co-based HAs alloys, like $Co_2MnSi$ and $Co_2CrAl$; where the presence of $Cr_{Co}$ and $Mn_{Co}$ disorders lead to the antiferromagnetic coupling between the doped atom and ideal atoms for $Co_2MnSi$ and $Co_2CrAl$ alloy, respectively [51,56]. Thus, creating $Y_X$ antisite disorder (within $X_2YZ$ HAs) could be an alternative promising way to make HMFi, and its further experimental investigations could be interesting.

Also, the present calculations of the total magnetic moment and the spin polarization of different disordered structures of $Co_2MnAl$ alloy (Section 3.2.2 & 3.2.3) collectively suggest that there is no straightforward rule that connects the electronic and magnetic properties to the disorder, rather it depends on the disorder-induced specific changes in electronic structure, which is different for different Heusler alloy as they have different electronic structures.

## 3.3 Effect of lattice distortions on structural, electronic, and magnetic properties of $Co_2MnAl$

In this next section, we present DFT results on the effect of lattice distortions by considering uniform compressive strain, uniform tensile strain, and tetragonal distortions on structural, electronic, and magnetic properties of $Co_2MnAl$ using 4 *f.u.* unit-cell (*i.e.*, 16 atom supercell). For modeling the uniform strain, the lattice parameter ($a$) is uniformly contracted and expanded by varying the volume of cubic unit-cell from $V_0$-10%$V_0$ to $V_0$+10%$V_0$, in 21 steps with step size of 1%$V_0$. Here, $V_0$ represents the volume corresponding to the fully relaxed cubic structure (or $L2_1$ ordered structure) and is calculated as $V_0 = (a_0)^3$; where $a_0$ is the optimized lattice parameter for relaxed $Co_2MnAl$ (5.69 Å). The uniform strain is defined as percent change in optimized volume *i.e.,* $\Delta V/V_0$ (× 100%) with $\Delta V = (V\text{-}V_0)$. This uniform strain leads to lattice parameters ranging from 5.50 Å to 5.88 Å. For tetragonal distortions, three different volumes for the tetragonal unit-cell are considered (*viz.*, $V_0$ and $V_0\pm5\%V_0$). The tetragonal distortion is then modeled at each volume with cell parameters, $a = b$ and $0.5 \leq c/a \leq 1.5$. For the tetragonally distorted cell with volume $V_0$, lattice parameters ranging between 4.97 Å $\leq a \leq$ 7.17 Å and 3.59 Å $\leq c \leq$ 7.46 Å, whereas for the tetragonal distorted cell with volume $V_0$-5% ($V_0$+5%), lattice parameter changes between 4.88 Å $\leq a \leq$ 7.05 Å (5.05 Å $\leq a \leq$ 7.29 Å) and 3.52 Å $\leq c \leq$ 7.33 Å (3.64 Å $\leq c \leq$ 7.58 Å). Tables S1, S2, S3 & S4 (in the supplementary material) depict the lattice parameters for the cubic strained and tetragonally distorted structures.

The range for lattice distortion, as mentioned above, is considered using the following factors: the experimental lattice parameter for cubic HAs is generally 1%-2% larger than the theoretically calculated lattice parameter, resulting from the heat treatment during the experiments or stress from growth technique such as sputtering [22]. Moreover, within the HA heterostructures, the lattice mismatch with the adjacent layers generally leads to a change in the lattice parameter of HA by 2%-3%. These alternations in lattice parameter are approximately equivalent to a maximum volume change of approximately ±10% in the optimized value. Therefore, volume change of ±10% has been considered by varying the unit-cell volume uniformly to model uniform strain.

For modeling the tetragonal distortion, three different volume values - $V_0$, ($V_0$+5%$V_0$) and ($V_0$-5%$V_0$) is considered for the tetragonal unit-cell volume, and then at each of these volumes, the *c/a* value is varied from 0.5 to 1.5 for modeling the tetragonal unit-cells. The choice of the tetragonal unit cell volume as volume of the $L2_1$-ordered structure ($V_0$) or with slightly varied volume ($V_0\pm5\%V_0$) is based upon the observation that *most of the technologically* important full cubic HAs with stable- or metastable- tetragonal phase often exhibit the same volume as in their cubic phase or show minor changes. For example - In the tetragonally distorted Co-, Fe- and Mn-based full Heusler alloy with high perpendicular magnetic anisotropy, the volume change is the order of 1% - 5% (with $1.2 \leq c/a \leq 1.4$) between stable cubic and tetragonal phase [57,58]; numerous all-*d*-metal based HAs (particularly Mn-Ni-V based) exhibit stable tetragonal structure with same equilibrium volume as of their cubic structure with $1.42 \leq c/a \leq 1.44$; and also in many ferromagnetic shape memory alloys, the volume for equilibrium state does not change under tetragonal distortion [59]. Meanwhile, the extremes of these distortions are chosen at relatively high values to study the trends in the results. In the following subsections, effect of uniform strain and tetragonal distortions would be discussed in detail.

## 3.3.1 Effect of uniform strain on structural, electronic, and magnetic properties of $Co_2MnAl$

The relative formation energy for the lattice distorted structure *w.r.t.* $L2_1$ ordered structure have been calculated, like for the point-defected structures in Section 3.2.1. In the case of lattice distortions (for uniform strains and tetragonal distortions), the last term in equation (1) does not exist, as the total number of atoms in the lattice distorted structures are the same as that of the ideal $L2_1$ structure. Accordingly, the RFE for the lattice distorted structures, or sometimes termed as the distortion formation energy, is given by

$$E_{form}(A) = E(A) - E(O) \tag{3}$$

where $E_{form}(A)$, E(A), and E(O) is the RFE of the lattice distorted structure, total energies of the lattice distorted structure (A), and total energies of the equivalent relaxed cubic structure ($L2_1$), respectively. The effect of uniform cubic strain (or *a* variation) on the structure, electronic, and magnetic properties are summarized in Figure 5. All uniform strained structures have positive RFE, as Figure 5(a) depicts. With increasing the strain values ($\Delta V/V_0$), the RFE increases in a parabolic way; emerging from the change in the potential energy of the strained structures.

The positive RFE for the uniformly strained structures suggests that the growth of uniform strained structure is less favorable during the formation of $Co_2MnAl$ alloy, compared to the $L2_1$ ordered structure. However, for a considerably wide range of cubic strain, corresponds to $-5\% \leq \Delta V/V_0 \leq 6\%$, the RFE for the uniform strained structure is less than the 0.1 eV/*f.u.* This is comparable to the thermal energy typically encountered during experiments and the criteria for distortion formation energy for most of the experimentally observed strains in full-HAs. Thus, it can be inferred that during the growth of $Co_2MnAl$; the uniform (or isotropic) strained structure with volume $0.95V_0 \leq V \leq 1.06V_0$ (equivalent to the lattice parameters between 5.60 Å to 5.81 Å) should be easily formed in experiments. Beyond that, the occurrence probability for the strained structure decreases rapidly with increasing strain values ($\Delta V/V_0$), due to the sharp increase in distortion formation energy.

Thereafter, the effect of strain on electronic properties was investigated using DOS. The spin polarized DOS plots under uniform strain are shown in Figure 6, where we present results only for few strain values, *viz;* $\Delta V/V_0 = 0, \pm 5\%$ and $\pm 10\%$. The Fermi energy is shifted to zero in Figure 6. Under the uniform strain, there were minor changes in the minority spin DOS near $E_F$, while the majority spin DOS remains nearly the same as for the relaxed structure. The general shape of DOS for the uniformly strained structure did not change, due to the similar crystallography as for the relaxed structure. For the negative strain ($\Delta V/V_0 \leq 0$), the smaller unit-cell volume of strained structure results in the broader and shallower bands because of the increased electronic orbitals overlapping, as illustrated in Figures 6(a) & 6(b). The opposite effect is observed with positive strain ($\Delta V/V_0 \geq 0$), and strained structure exhibits narrower and peakier bands, as seen in Figures 6(c) & 6(d). Apart from these minute changes in DOS shape, some other minor changes in DOS for uniformaly strained structures include: change in the minority energy gap width ($E_{gap} = E_c\text{-}E_v$) and shifting in Fermi level (or equivalently change in the gap center $\{(E_c+E_v)/2\}$ or shifting of minority energy band *w.r.t.* $E_F$). All these changes are somewhat similar to what was observed in point defect cases, such as all changes happening in the minority DOS only. As seen from Figure 5(d), $E_{gap}$ decreased with increasing the cell volume and changed from 0.67 eV to 0.57 eV at extreme of strain [60]. As for the free-electron gas, the $E_F$ changes with unit-cell volume as $\mathbf{E}_{\boldsymbol{F}}\ \boldsymbol{\alpha}\ \boldsymbol{V}^{-2/3}$; therefore, a change in $E_F$ with cell volume is also expected for strained structures (Figure S4(a), supplementary material). Other factors influencing $E_F$ and $E_{gap}$ under strain include changes in the electronic localization and exchange coupling, as depicted in PDOS plots and are also explained for $Co_2MnGe$, $Co_2CrAl$, and $Co_2FeAl$ alloys in literature [14,23,24].

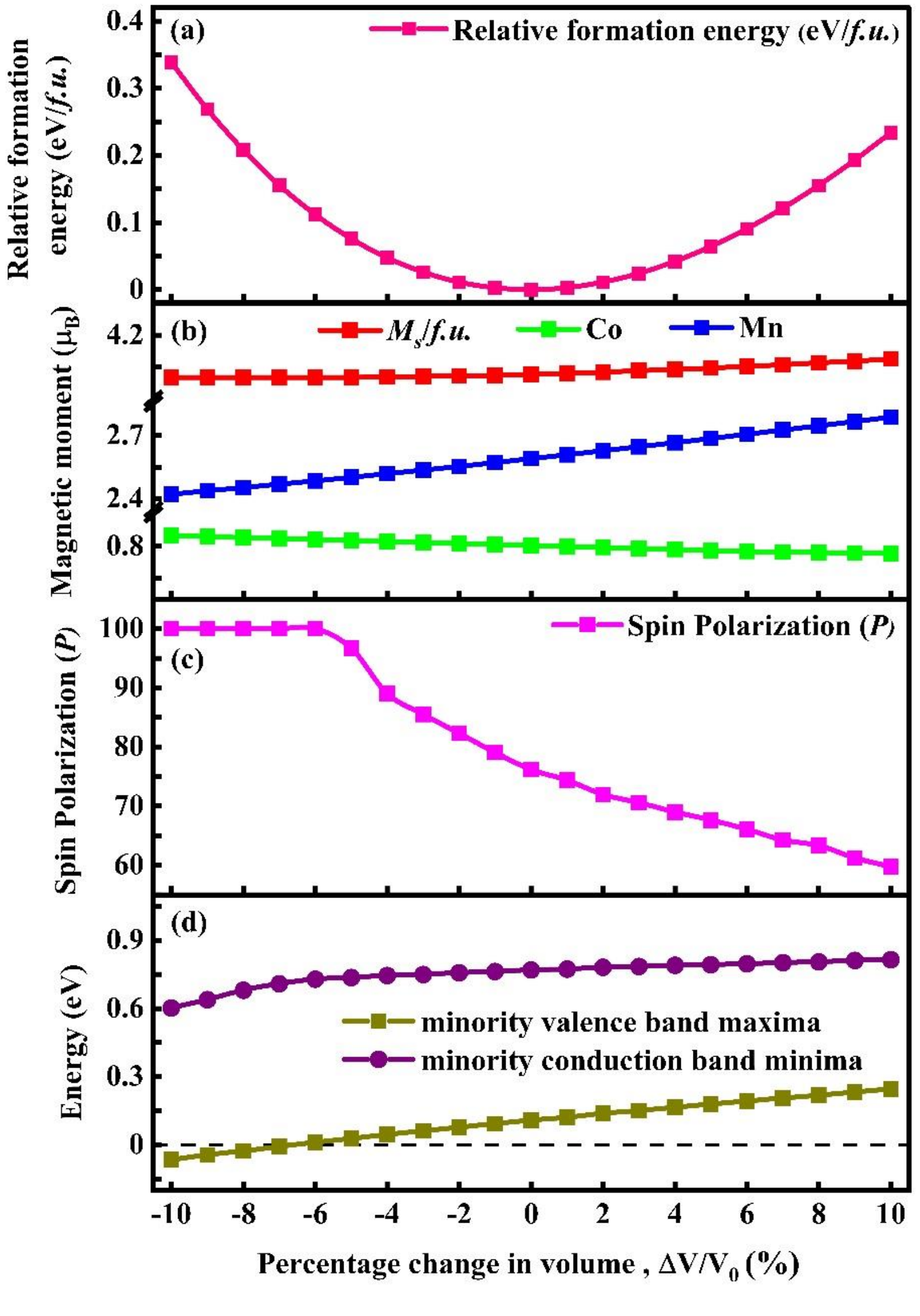

**Figure 5:** (a) Relative formation energy (b) total and atomic magnetic moments (c) spin polarization and (d) minority valence band maxima and minority conduction band minima for uniform (or isotropic) strained $Co_2MnAl$ alloy as a function of percent change in unit-cell volume ($\Delta V/V_0 = \{V\text{-}V_0\}/V_0$). The horizontal short-dashed line in Figure 5 (d) indicates the Fermi level.

The trends in minority energy gap, $P$, atomic magnetic moments and total magnetic moments, for the cubic strained structures are given in Figures 5(b) - 5(d). For negative strains (or uniform compressive strain), $Co_2MnAl$ possess enhanced *P* along with 100% polarization for the strain values upto $\Delta V/V_0 \leq -6\%$. Above that, a monotonic decrease in *P* is observed with increasing strain value (*i.e.,* with increasing unit-cell volume) and drops to ~ 60% at $\Delta V/V_0 = 10\%$. The primary reason for the change in *P* is the shifting of $E_F$ in the minority spin channel. The $E_F$ is pinned in the valley in the majority spin channel, and the shifting occurs only independently in the minority spin channel, like in the case of point defects. At the extreme negative strain, $E_F$ is situated in the energy gap, resulting in perfect half-metallic nature for the strained structure. With increasing positive strain, $E_F$ shifts towards the minority valence band. As a result, for $(\Delta V/V_0) \geq -6$ %, *P* decreases *linearly* for uniform strained structures while increasing the cell volume.

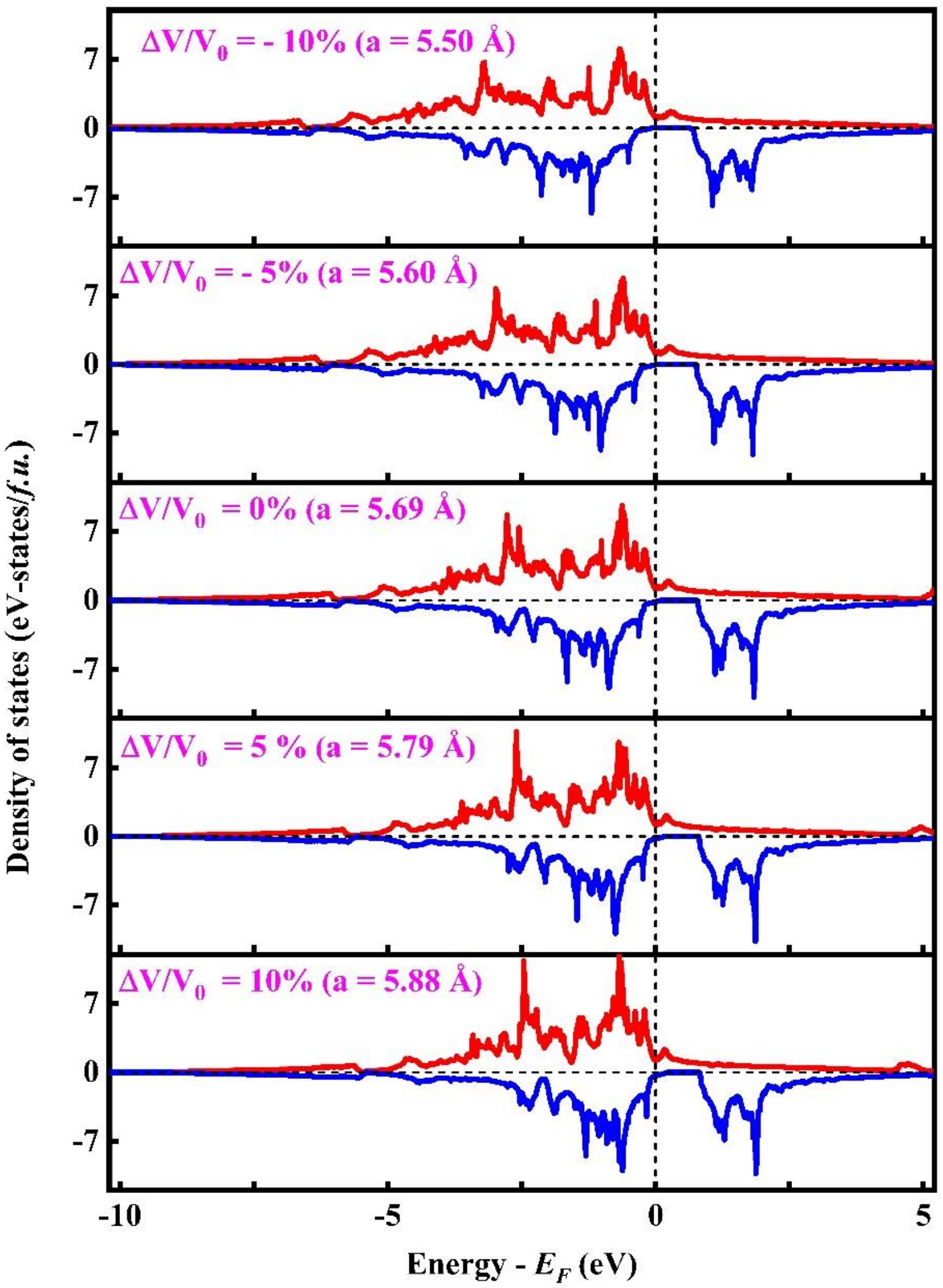

**Figure 6**: The DOS plots for −10%, −5%, 0%, 5%, and 10% uniform strained ($\Delta V_0/V_0$) $Co_2MnAl$ alloy. The red and blue solid lines show the majority and minority DOS.

The total magnetic moment remains nearly constant for all investigated strain values. Specifically, for strain values upto $\Delta V/V_0 \leq -6\%$, $M_s$ is an integer (*i.e.*, 4.00 $\mu_B$/*f.u.*). This is consistent with Slater Pauling rule for half-metallic full HAs (*i.e.*, integer magnetic moments for HAs with 100% *P*). For other strain values ($\Delta V/V_0 > -6\%$), changes in $M_s$ were observed at second decimal points. Concerning the AMM; when the lattice expands; Co and Mn atoms go towards a more isolated, atomic-like situation, due to the increased localization (or because of reduced hybridization with the Co-*d* states). This increased localization for Co- and Mn-atoms with increasing cell volume is also evident in their PDOS, which is shown by the enhanced peaky and narrow DOS structure (Figure S5, supplementary material). As a result, due to increased localization, Mn magnetic moments increases with increase in the cell volume [61,62]. In contrast, Co AMM shows the opposite trend and decreases with positive strain. However, the rate of change for Co AMM *w.r.t.* strain value is very small compared to the rate of change for Mn AMM, which is attributed to itinerant nature of Co atom than the Mn atom. Thus, the increment in Mn AMM is compensated for the Co AMM and interstitial moments; leaving the total spin moment constant as in the case of ideal structure across all strain values. Generally, the change in Co AMM in Co-based HAs with strain predominantly depends upon the specific Y-type atom (in $X_2YZ$ HAs); whereas for Y-type atom, AMM increase with increasing uniform strain. For example, in the $Co_2MnSi$ and $Co_2CrAl$ alloys, Co AMM decreases with increasing positive strain; while it increases in $Co_2FeAl$ alloy. Therefore, our finding regarding the trend in the change in AMMs are consistent with that of other Co-based HAs. In summary, it can be deduced that uniform strain has minimal impact on $M_s$, while *P* changes significantly, depending upon the strain value.

### 3.3.2 Effect of tetragonal distortions on structural, electronic, and magnetic properties of $Co_2MnAl$

As mentioned in the introduction section, tetragonal distortion is anticipated in thin films and is helpful for many practical technological applications. Therefore, this section will discuss the effect of tetragonal distortion on the structural, electronic, and magnetic properties of $Co_2MnAl$ alloy. Figures 7 to 10 present their structural, electronic, and magnetic properties of tetragonally distorted structures, at different *c/a* values, with **$V = V_0$, ($V_0$ -5%$V_0$)**, and ($V_0$+5%$V_0$) volumes, respectively. Based on the calculated RFE as presented in Figures 7(a), 9(a) & 10(a); it is evident that the tetragonally deformed structures have very high positive RFE, which increase with *c/a* values. Therefore, the tetragonally distorted structures are less likely to be formed (or relatively less favored for formation) compared to the *L2₁*-ordered structure.

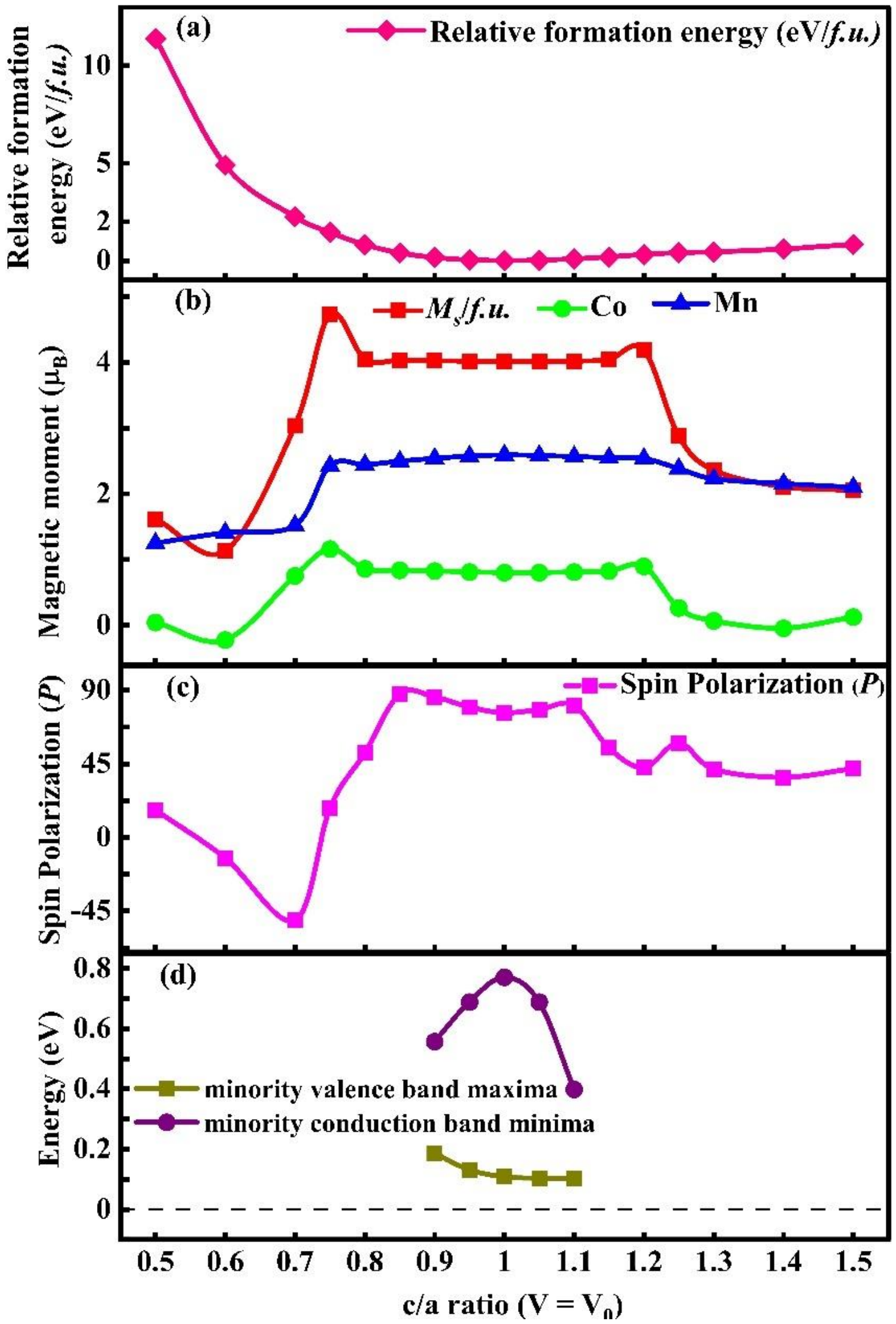

**Figure 7:** (a) Relative formation energy (b) total and atomic magnetic moments (c) spin polarization and (d) minority-spin valence band maxima and conduction band minima as a function of *c/a* value for tetragonal structures with $V_0$ unit-cell volume. The horizontal short-dashed line in Figure 7(d) indicates the Fermi level.

However, *the tetragonally distorted structures with $V_0$ volume*, for very small distortion values ranging between *c/a* = 0.95 to *c/a* = 1.10, have low RFE ≤ 0.1 eV/*f.u.*; making their occurrence highly probable. Beyond these *c/a* values; RFE increases rapidally with increasing |Δ*c/a*| values, making their occurrence probability decreasing with increasing |Δ*c/a*| values. As seen from Figures 9(a) & 10(a), a similar trend is observed for *tetragonally distorted structures with ($V_0 \pm 5\%V_0$) volumes*, except for the *c/a* values with RFE ≤ 0.1 eV/*f.u*. For the *tetragonally distorted structures with ($V_0+5\%V_0$) volumes*, this *c/a* range changes between 0.95-1.10; whereas for the *tetragonally distorted structures with ($V_0+5\%V_0$) volumes* have RFE ≥ 0.1 eV/*f.u.* for all considered *c/a* values.

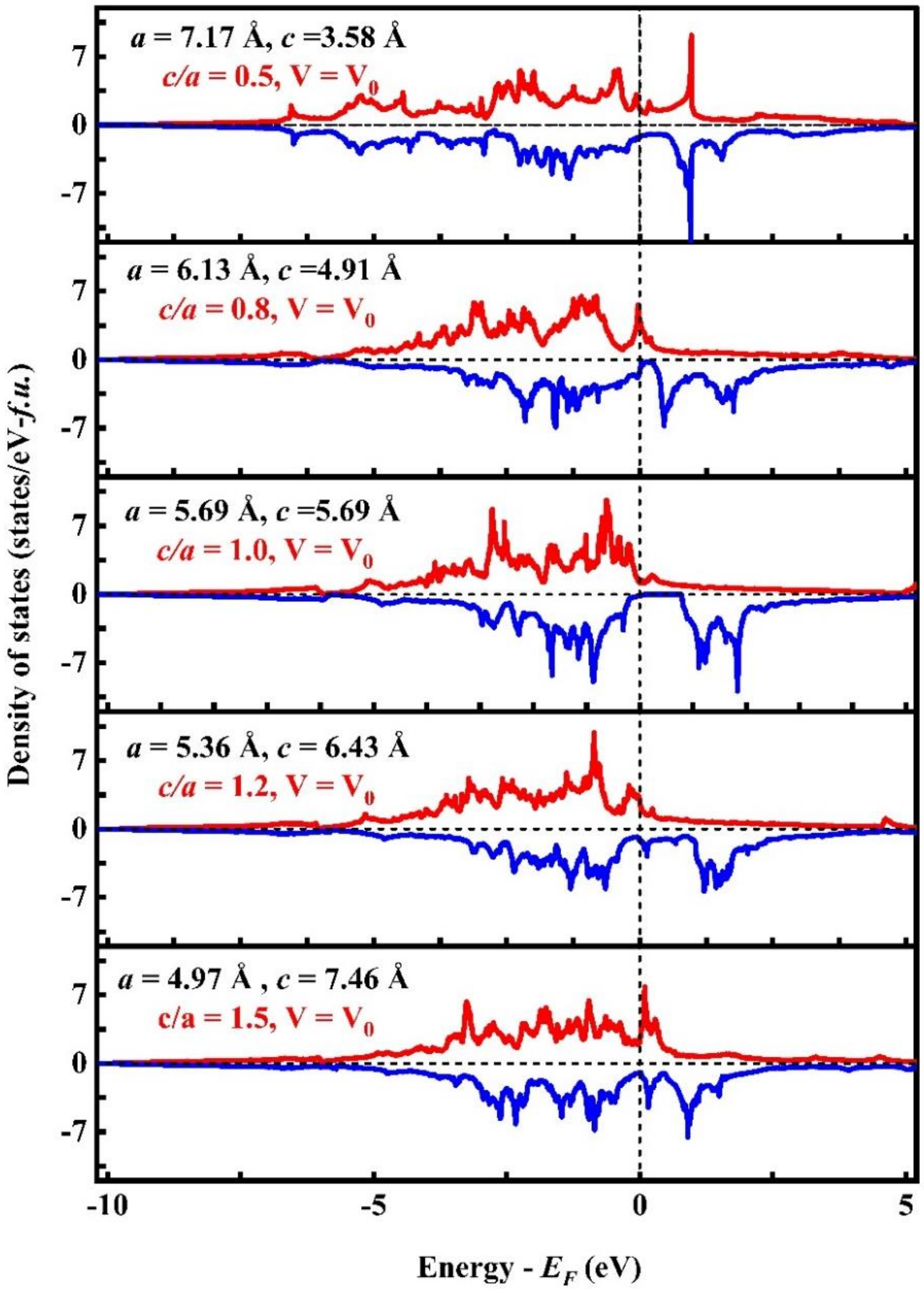

**Figure 8 :** DOS plots for tetragonally distorted structures with *c/a* = 0.5, 0.8, 1.0, 1.2 and 1.5 with $V_0$ unit-cell volume. The red and blue solid lines show the majority and minority DOS.

Also, for all *c/a* values, *tetragonal structures with compressed volume (i.e., $V=V_0-5\%V_0$)* have the highest RFE, followed by *the tetragonal distorted structure with increased volume* (*i.e., $V = V_0+5\%V_0$).*In contrast, those with $V_0$ volume have the lowest REF. As a result, $|\Delta c/a|$ with RFE $\leq$ 0.1 eV/*f.u.* also varies *w.r.t.* tetragonal unit-cell volume. At a particular *c/a* value, the change in the RFE *w.r.t.* tetragonal unit-cell volume can be attributed to the change in potential energy of the tetragonal structure *w.r.t.* unit-cell volume. Therefore, among the $V_0$ and ($V_0\pm5\%$) volumes, the tetragonal distortion is most likely to occur with the $V_0$ volume. This also explains why most full HAs have identical volumes in the tetragonal and cubic phases. Another significant point to note about the RFE of the tetragonally distorted structures is that, RFE is much larger for the compression along the *c*-axis (with elongation of the *ab*-plane) compared to the elongation along the *c*-axis (with compression of the *ab*-plane), regardless of their volume. Thus, it is concluded that the favored condition for tetragonal distortion is elongation along the *c*-axis (with compression of the *ab*-plane).

Following that, let discuss about the DOS and *P* for the tetragonal distorted structures. For the tetragonally distorted structures, DOS undergoes significant alterations due to the weight redistribution resulting from the reduced symmetry of the crystal structure. Consequently, the principal features of HAs band structure, *i.e.*, peak and valley characteristics disappear; and a broader and shallow DOS , compared to the optimized cubic structure, is observed for the tetragonally distorted structures. This phenomenon is easily noticeable in their DOS plot as depicted in Figure 8, which shown the DOS plot for the tetragonal structures with $\mathbf{V_0}$ volume and different *c/a* values. For the tetragonally distorted structures, as far as the *c/a* value deviates from 1.0, the DOS becomes more smoothly distributed and dispersive. However, the DOS remained relatively unaltered for minimal distortion values (*i.e.,* for very small $|\Delta c/a|$ values, around 1.0), depending upon deformed unit-cell volume.

For the *tetragonally deformed structures with $V_0$ volume,* with very small distortion values ($0.85 \leq c/a \leq 1.10$); DOS shape nearly the same as in case of the *$L2_1$* ordered structure with intact SP valley (not shown). This leads to a nearly half-metallic behavior for them, along with very high spin polarization like *$L2_1$*-ordered structure. Further large distortion (*i.e.,* higher $|\Delta c/a|$ values ) results in the closing of the SP valley due to the large overlapping of the atomic orbitals along the contraction direction. Therefore, a dominating metallic nature with a considerable reduction in *P* is observed for the remaining deformed structures (*i.e.,* with $c/a \leq 0.85$ and $c/a \geq 1.10$) as seen in Figure 7(c). Specially, as *c/a* values increased from 1.10, *P* declined almost monotonically, reaching 42% at $c/a = 1.5$. Conversely, when *c/a* value decreased from 0.85, *P* showed a random drop and it even changed its sign at $c/a = 0.7$ and 0.6 with *P* values of -50% and -12%, respectively. Here, the primary influence on *P* for the tetragonally distorted structures arises from the altered shape of DOS. Unlike the case of the uniformly strained structures, the shift in the Fermi level is not as significant as for the tetragonally distorted structures (see Figure S4(b), supplementary material). This is attributed to the simultaneous elongation and contraction while maintaining the constant unit-cell volume in the tetragonally distorted unit-cell. Therefore, the shifting of $E_F$ has negligible contribution in the alternation in $P$; like in the case of uniform strain. *For the tetragonal distorted structures with altered unit-cell volume ($V_0$-5%$V_0$ and $V_0$+5%$V_0$)*, DOS as well as $P$ follow the nearly same pattern as for the tetragonally structures with $V_0$ volume. However, SP valley or the minority spin gap was found to be presented over a slightly wider range of *c/a* (from 0.80 to 1.15) for the tetragonally distorted cell with "decreased volume" as compared to structures with "increased volume" (from 0.90 to 1.10), and structures with "optimized volume" (from 0.85 to 1.10) as indicated by Figure 7(d), 9(b) & 10(d).

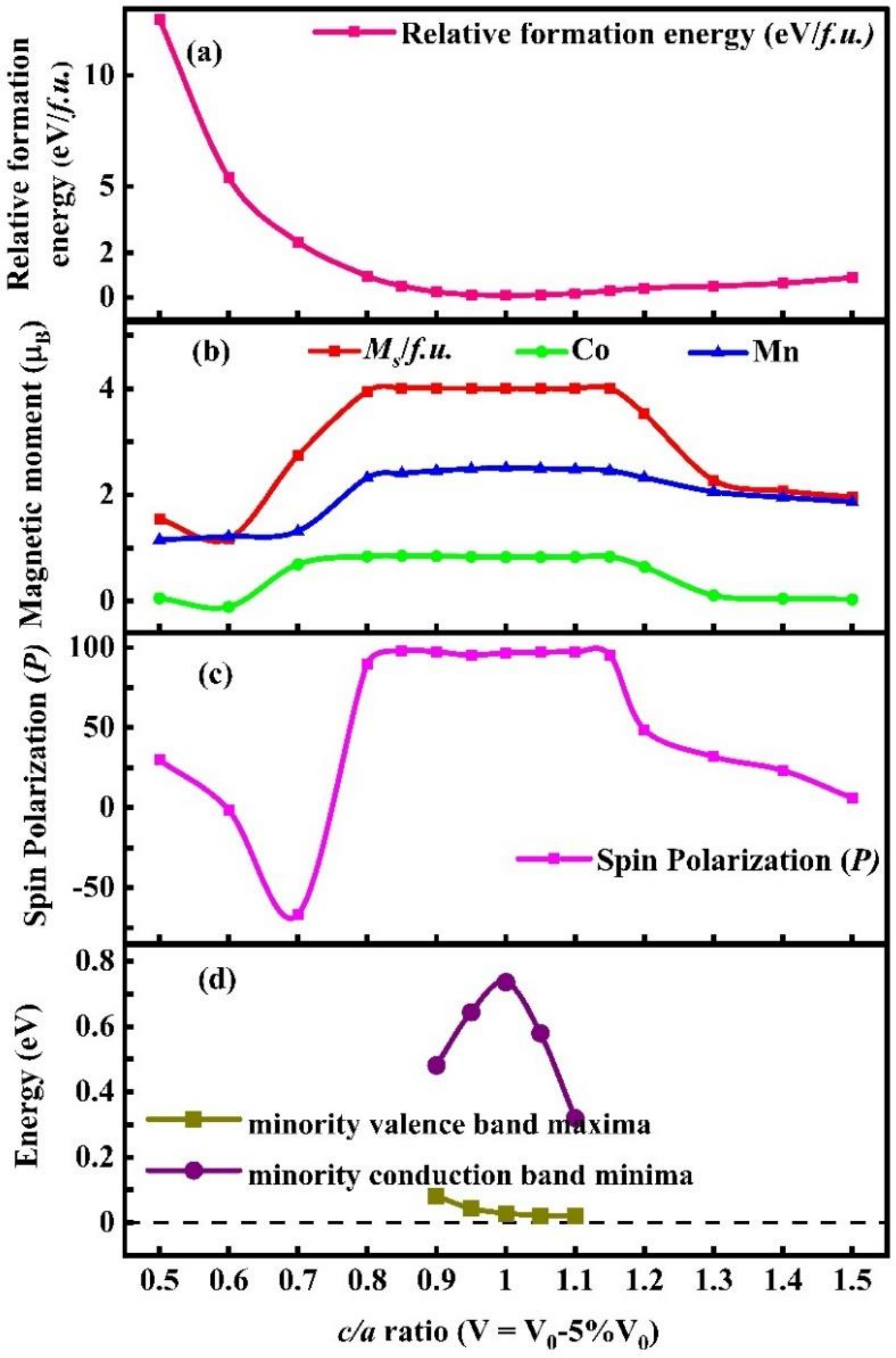

**Figure 9:** Same as Figure 7, but here for the tetragonal structures with ($V_0$-5%$V_0$) unit-cell volume.

Furthermore, at a particular *c/a* value, the distorted structures with "decreased volume ($V_0$-5%$V_0$)" maintain the high spin polarization values compared to distorted structures having $V_0$ and ($V_0$+5%$V_0$) volume as seen in Figures 7(c), 9(c) & 10(c). The structures with ($V_0$+5%$V_0$) volume have the lowest *P*. For the tetragonally deformed structures, at a particular *c/a* value, the decline in *P* with increasing the unit-cell volume (from $V_0$-5%$V_0$ to $V_0$+5%$V_0$) can be attributed to the weakening covalent hybridization on increasing the cell volume [24]. Thus for *P* of tetrgonally distorted cells, it can be summarized that *P* changes arbitrarily and shows the non-monotonic behavior with *c/a* values, which primarily originates from the altered shape of DOS including the closing of the minority energy gap or SP valley. Also, for small distortion, the deformed structures maintain high *P* and the minority spin gap, and the large distortion (large |Δ*c/a*| value) results in reduced *P*. For a particular *c/a* value, *P* decreased on increasing the cell volume.

The total magnetic moment also follows a similar trend as *P*, *i.e.*, for the small deviation of the *c/a* value from 1.0, distorted structures have nearly the same moment values as of $L2_1$-ordered structure; whereas the large deviations of *c/a* from 1.0 lead to a significant alternation in magnetic moment values for the tetragonally distorted structures (Figures 7(b), 9(b) & 10(b)). For 0.80 ≤ *c/a* ≤ 1.15, magnetic moment values are nearly same as in the case of relaxed cubic structure (~ 4 $\mu_B$/*f.u.*) *for the tetragonally distorted structure with* $V_0$ *volume*. The corresponding ranges for ($V_0$-5%$V_0$) *and* ($V_0$+5%$V_0$) *volume distorted cells* are $0.85 \leq c/a \leq 1.15$ and $0.90 \leq c/a \leq 1.10$, respectively. For further large change in *c/a* values, change in hybridization primarily resulting from the change in nearest neighbors, causes a significantly different magnetic moment values, irrespective to their volume. In addition to Figures 7, 9 and 10; RFE, magnetic moments, and spin polarization *for tetragonal distorted structures with* $V_0$ *and* ($V_0$±5%$V_0$) *volumes* are also tabulated in Tables S4, S5 & S6 in the supplementary material for better comprehension.

Finally, we discuss about the structural stability for the tetragonally distorted $Co_2MnAl$ alloy. Recall that, the RFE also provides information relative thermodynamic stability of the defected structures, in addition to the ease of occurrence of the disorder. As the tetragonally distorted structures have very high positive RFE as observed from Figures 7(a), 9(a) & 10(a); therefore, the tetragonally distorted structures for $Co_2MnAl$ HA are relatively less stable than the ordered cubic structure. The reason behind the instability of the tetragonally distorted $Co_2MnAl$ lies in its electronic configurations and can be attributed to the higher valence band energies for the tetragonal structures compared to the relaxed cubic structure. The valence band energy for the relaxed cubic structure or the tetragonally distorted structure, can be written as $E_{band} = \int_{E=Ev}^{EF} \mathrm{d}E\, DOS(E) * E$, as discussed in Ref. [26] . Compared to the relaxed cubic structure, the tetragonally deformed structures have higher DOS near the $E_F$ (see Figures 6 & 8). Hence, for the tetragonally distorted structure, in the conjugation with number of valence electrons $N_v = \int_{E=E_v}^{E_F} dE\, DOS(E)$; the increased DOS around $E_F$, lead to the higher band energies for the tetragonal structures. However, the absence of metastable or stable tetragonal phase for $Co_2MnAl$ in this study may be limited due to the constrained volume approach and/or the shuffling of atoms in larger and big supercell for making tetragonal cell, and therefore require further study [63].

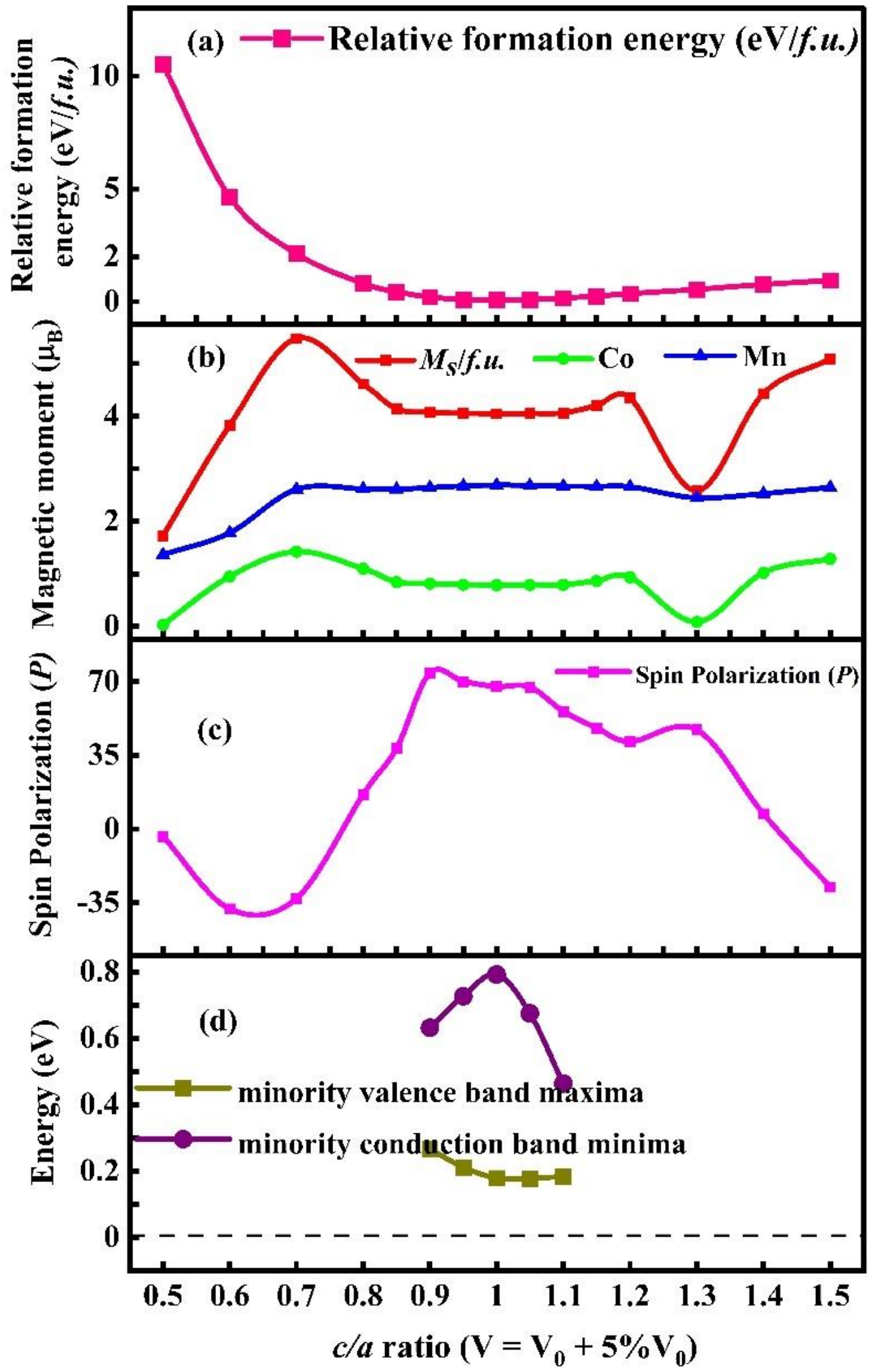

**Figure 10:** Same as Figure 7, but here for the tetragonal structures with ($V_0$+5%$V_0$) unit-cell volume.

# 4. Conclusion

The signature of structural imperfections is frequently observed in the Heusler alloys like a perturbation to the ordered structure, which strongly affects their physical properties. Given that the $Co_2MnAl$ have been reported to have some exceptional properties in presence of point defects and lattice distortions, such as – appearance of SGS nature, emergence of PMA and enhanced spin polarization; along with partial availability of studies for the effect of point defects and lattice distortions, a detailed study for the effect of various point defects (binary antisite disorder, ternary antisite disorder, vacancy defects) and the lattice distortions (uniform cubic strains and tetragonal distortions) on the structural, electronic and magnetic properties of the technologically important Heusler alloy $Co_2MnAl$ have been done explicitly for uncovering some remarkable results. While the point defect is considered up to a maximum disorder concentration of 12.50% using 64 atoms supercell, the cubic strain ($\Delta V/V_0$, %) is simulated within the range of ±10% (corresponding to a variation in the lattice parameter ranging from 5.50 Å to 5.88 Å) and the tetragonal deformation is modelled by varying *c/a* value between 0.5-1.5 at three different unit-cell volumes ($V_0$ and $V_0 \pm 5\% V_0$) using 16 atoms supercell.

The Mn deficient structures resulting from $Al_{Mn}$ binary antisite disorder ($Co_2Mn_{0.9375}Al_{1.0625}$ and $Co_2Mn_{0.875}Al_{1.125}$) and ($Co_{Mn}$+$Al_{Mn}$) ternary antisite disorder ($Co_{2.0625}Mn_{0.875}Al_{1.0625}$), exhibit negative relative formation energies *w.r.t.* $L2_1$-ordered structure; due to which they could spontaneously occur during the $Co_2MnAl$ growth. The $Co_{Al}$ antisite-disordered structures ($Co_{2.0625}MnAl_{0.9375}$ and $Co_{2.125}MnAl_{0.875}$), the ($Co_{Al}$+$Mn_{Al}$) antisite-disordered structure ($Co_{2.0625}Mn_{1.0625}Al_{0.875}$), and mono- and di-vacancies defected structures for Co-, Mn- and Al-atomic sites have very high positive relative formation energies *w.r.t.* $L2_1$-ordered structure, which make their occurrence unlikely or expected to have very small density. The $Co_{Al}$ binary antisite disordered structures ($Co_{2.0625}MnAl_{0.9375}$ and $Co_{2.125}MnAl_{0.875}$), $Mn_{Al}$ binary antisite disorder structures ($Co_2Mn_{1.0625}Al_{0.9375}$ and $Co_2Mn_{1.125}Al_{0.875}$) and ($Co_{Al}$+$Mn_{Al}$) ternary antisite disordered structures ($Co_{2.0625}Mn_{0.875}Al_{1.0625}$) exhibit perfect half-metallicity *i.e.*, 100% spin polarization and integer magnetic moment according to the Slater Pauling rule. The rest of the antisite disorders have marginal effect on half-metallic properties and the disordered structures maintain high spin polarization ( ≥ 70%) and nearly the same magnetization as in fully ordered $L2_1$ structure. Furthermore, emergence of the half-metallic ferrimagnetism phenomenon in the $Mn_{Co}$ binary antisite and ($Mn_{Co}$+$Al_{Co}$) ternary antisite disordered structures (namely $Co_{1.9375}Mn_{1.0625}Al$, $Co_{1.875}Mn_{1.125}Al$ and $Co_{1.875}Mn_{1.0625}Al_{1.0625}$), resulting from the antiparallel orientation between the atomic moments of the disordered atom and neighboring transition metal elements, suggest a possible method for synthesizing the half-metallic ferrimagnetic material and require further experimental investigations. For all considered antisite disordered structures (binary as well ternary), analyses of the partial density of states and atomic magnetic moments show that disordered (substituted) atoms solely owed the changes in the spin polarization and magnetization of the disordered $Co_2MnAl$ and the disorder's effect on electronic and magnetic properties is localized in nature. However, unlike the case of antisite disorders; the vacancy defects significantly affect the electronic and magnetic properties of $Co_2MnAl$. They were found to cause the narrowing of the pseudogap (Al-vacancies), including vanishing in some cases (Co- and Mn-vacancies). Another contrasting feature of vacancy defect is that the effect of defects survives up to several neighbors.

For the lattice distortions, the uniform strain ranging between $-5\% \leq \Delta V/V_0 \leq 6\%$ is most probable in the real sample due to their low relative formation energies *w.r.t.* $L2_1$-ordered structure. Due to the same crystal structure, the DOS remains unaltered under the uniform (isotropic) strain. However, the shifting of $E_F$ in the minority spin channel causes the appearance of half-metallicity for $-10\% \leq \Delta V/V_0 \leq -7\%$ (corresponding to 5.50 Å $\leq a \leq$ 5.56 Å). The spin polarization decreases monotonically with the increase in the positive strain, dropping to ~60% at $\Delta V/V_0 = +10\%$ ($a$ = 5.88 Å). Irrespective of the magnitude of the strain, the total magnetic moment remains unchanged under the uniform strain. On the other hand, for tetragonal distortion, expect for small range of *c/a* value (mostly 0.9-1.1); the tetragonally distorted structures have very high relative formation energies for all considered *c/a* values and unit-cell volumes ($V_0$ and $V_0 \pm 5\% V_0$). Therefore, tetragonal distortion is less probable to occur than the $L2_1$ ordered structure. Also, the elongation along the *c*-axis (with compression of the *ab*-plane) is the favored condition for the occurrence of tetragonal distortion than compression along the *c*-axis (with elongation of the *ab*-plane). The high spin polarization and minority spin pseudogap are found to be kept only for very small deviation of *c/a* value from 1.00; notably for $0.85 \leq c/a \leq 1.10$ at $V_0$ unit-cell volume, for $0.80 \leq c/a \leq 1.15$ at ($V_0$-5%$V_0$) unit-cell volume, for $0.90 \leq c/a \leq 1.10$ at ($V_0$+5%$V_0$) unit-cell volume; resulting high spin polarization and nearly same magnetization as of $L2_1$ ordered structure. However, more substantial deviations in *c/a* value (irrespective of the unit cell volume) lead to significant changes in the DOS-shape, including the pseudogap closing; resulting in the change in spin polarization and magnetic moment in an arbitrary way.

The present study shows that while some structural imperfections (binary antisite disorder, ternary antisite disorder and uniform strain) could be beneficial for $Co_2MnAl$ from the spintronics point-of-view, some other (vacancy defects and tetragonal distortion) could have adverse effects on the spintronics required properties, and therefore should be avoided by carefully selecting the substrates or insertion of spacer layers and by controlling the growth mechanism. We believe that the present results would be useful for the experimentalists to form thin films of high quality as desired in spintronic devices. We did not find any stable or metastable non-cubic structure for $Co_2MnAl$. The other possible volume of the tetragonally deformed cell and other possible high-symmetric lattice deformations like – octahedral distortion could be studied for stable or metastable non-cubic structure for $Co_2MnAl$. Also, the study of the disordered structures combining the antisite disorder and lattice distortions is a subject worthy of further investigation.

# ACKNOWLEDGMENTS

The authors would like to thank the IIT Delhi HPC facility for computational resources. A.K. acknowledges Council of Scientific and Industrial Research (CSIR) India, for the senior research fellowship. The authors thank Profs. M. Ali Haider, B. K. Mani, S. Bhattacharya, and R. S. Dhaka of IIT Delhi, and Prof. Manish Kumar, School of Physical Science, JNU Delhi for helpful discussions.

# <u>Supplementary Material</u>

## A. <u>GGA+U results for $Co_2MnAl$ alloy</u>

- **<u>Spin polarized DOS plot for $Co_2MnAl$ calculated with GGA and GGA+U method</u> :**

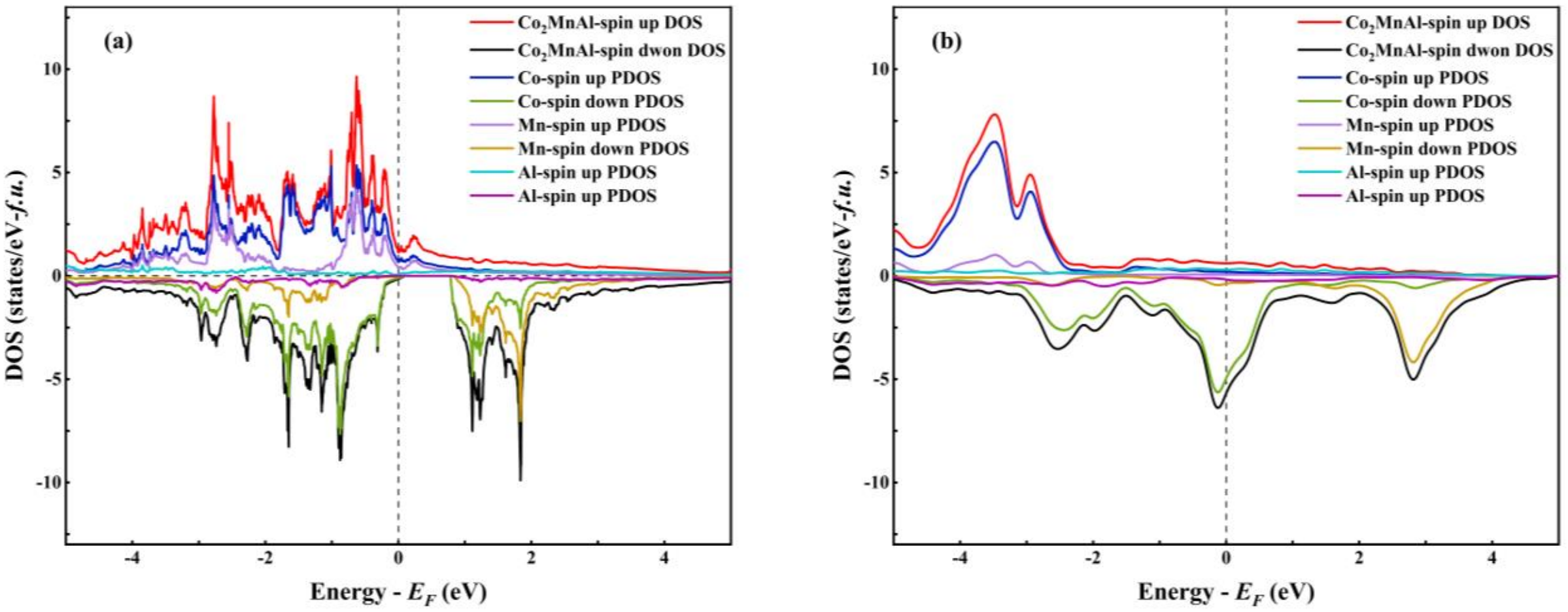

**Figure S1**: Total and atomic density of states of $Co_2MnAl$ alloy with (a) GGA and (b) GGA+U method.

## B. Effect of point defects on structural, electronic, and magnetic properties of $Co_2MnAl$ alloy:

- **Fermi energies for point defected structures:**

**Table S1:** The Fermi energy for binary antisite disordered structures and vacancy defected structure with different disorder degree(s), represented as a function of *x*. It should be noted that for $L2_1$-ordered structure, Fermi energy of $Co_2MnAl$ is 15.99 eV.

| Disorder type (stoichiometric formula) | Fermi energy (eV) *for* x = *0.0625* | Fermi energy (eV) *for* x = *0.125* |
|---|---|---|
| $Co_{Mn}$ antisite ($Co_{2+x}Mn_{1-x}Al$) | 16.02 | 16.03 |
| $Co_{Al}$ antisite ($Co_{2+x}MnAl_{1-x}$) | 16.12 | 16.24 |
| $Mn_{Co}$ antisite ($Co_{2-x}Mn_{1+x}Al$) | 16.00 | 15.98 |
| $Mn_{Al}$ antisite ($Co_2Mn_{1+x}Al_{1-x}$) | 16.13 | 16.28 |
| $Al_{Co}$ antisite ($Co_{2-x}MnAl_{1+x}$) | 15.82 | 15.62 |
| $Al_{Mn}$ antisite ($Co_2Mn_{1-x}Al_{1+x}$) | 15.90 | 15.78 |
| Co vacancy ($Co_{2-x}MnAl$) | 15.76 | 15.50 |
| Mn vacancy ($Co_2Mn_{1-x}Al$) | 15.75 | 15.50 |
| Al vacancy ($Co_2MnAl_{1-x}$) | 16.00 | 16.00 |

**Table S2:** The Fermi energy for ternary antisite disordered structures with different disorder degree(s), represented as a function of *x*.

| Disorder type (stoichiometric formula) | Structure nature | Fermi energy (eV) *for* x = *0.0625* |
|---|---|---|
| $Mn_{Co}$ + $Al_{Co}$ ($Co_{2-2x}Mn_{1+x}Al_{1+x}$) | Co-deficient | 15.80 |
| $Co_{Mn}$ + $Al_{Mn}$ ($Co_{2+x}Mn_{1-2x}Al_{1+x}$) | Mn-deficient | 15.91 |
| $Co_{Al}$ + $Mn_{Al}$ ($Co_{2+x}Mn_{1+x}Al_{1-2x}$) | Al-deficient | 16.26 |

# <u>Atomic magnetic moments for the point defected structures</u> :

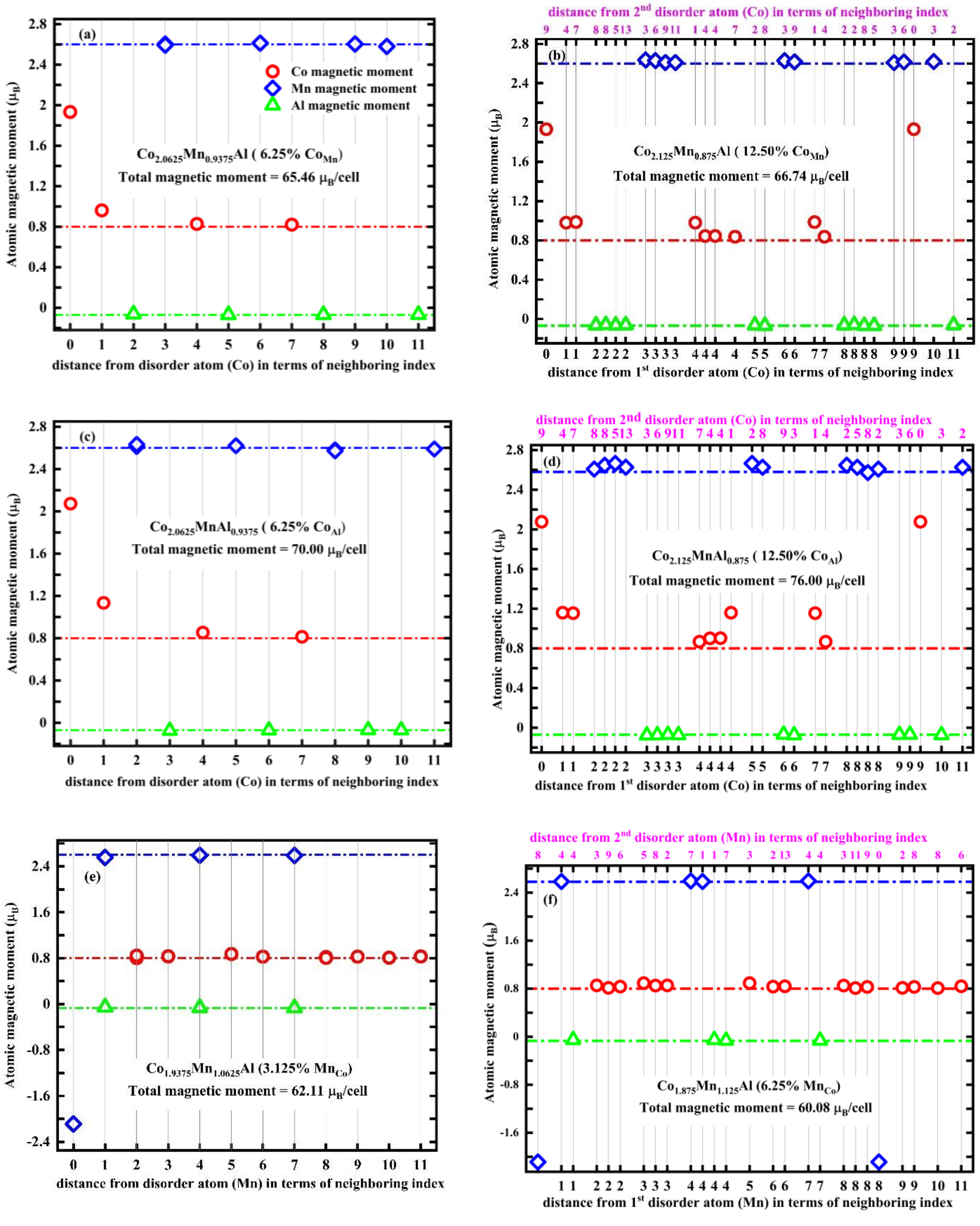

**The Figure continues on next page.…**

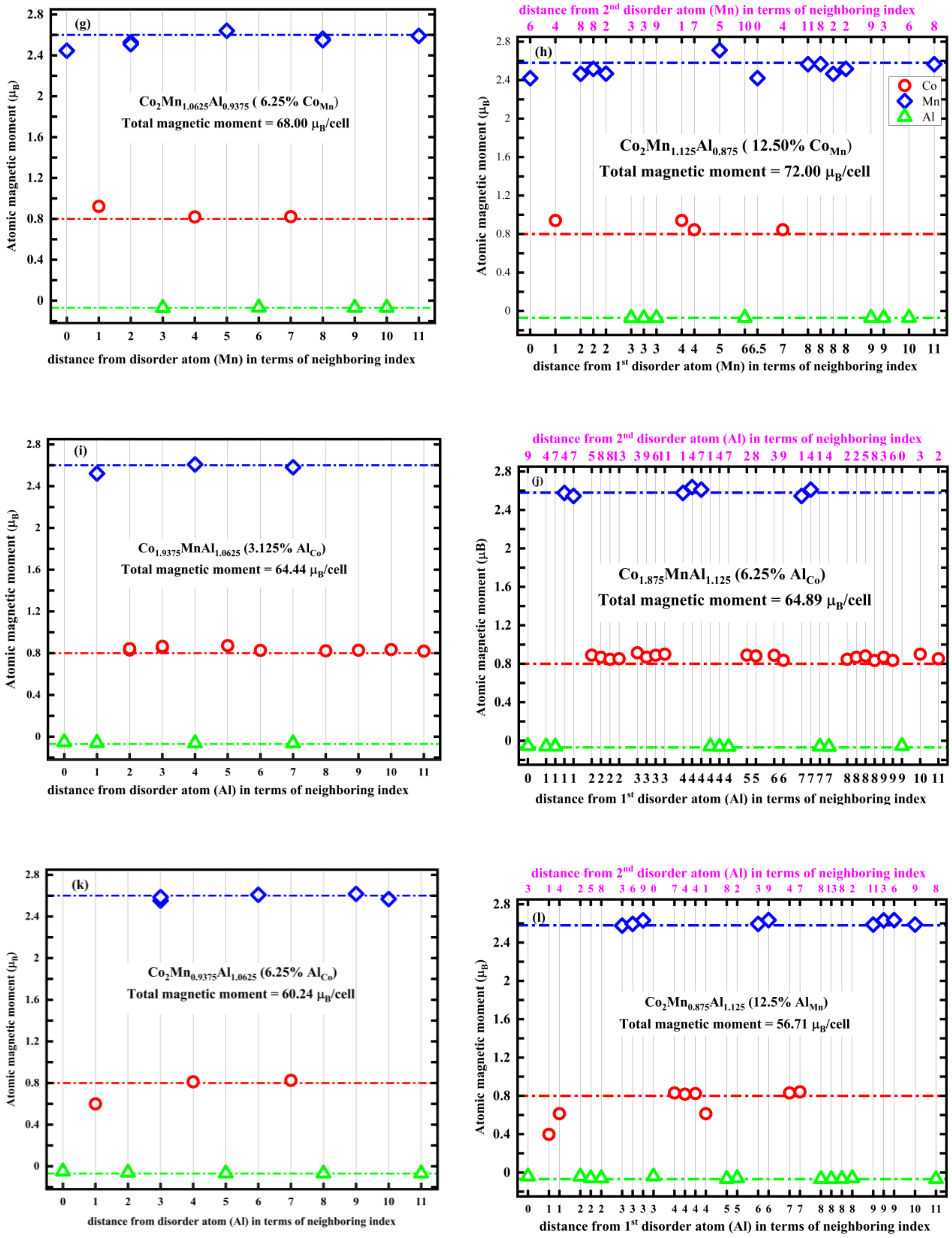

**The Figure continues on next page….**

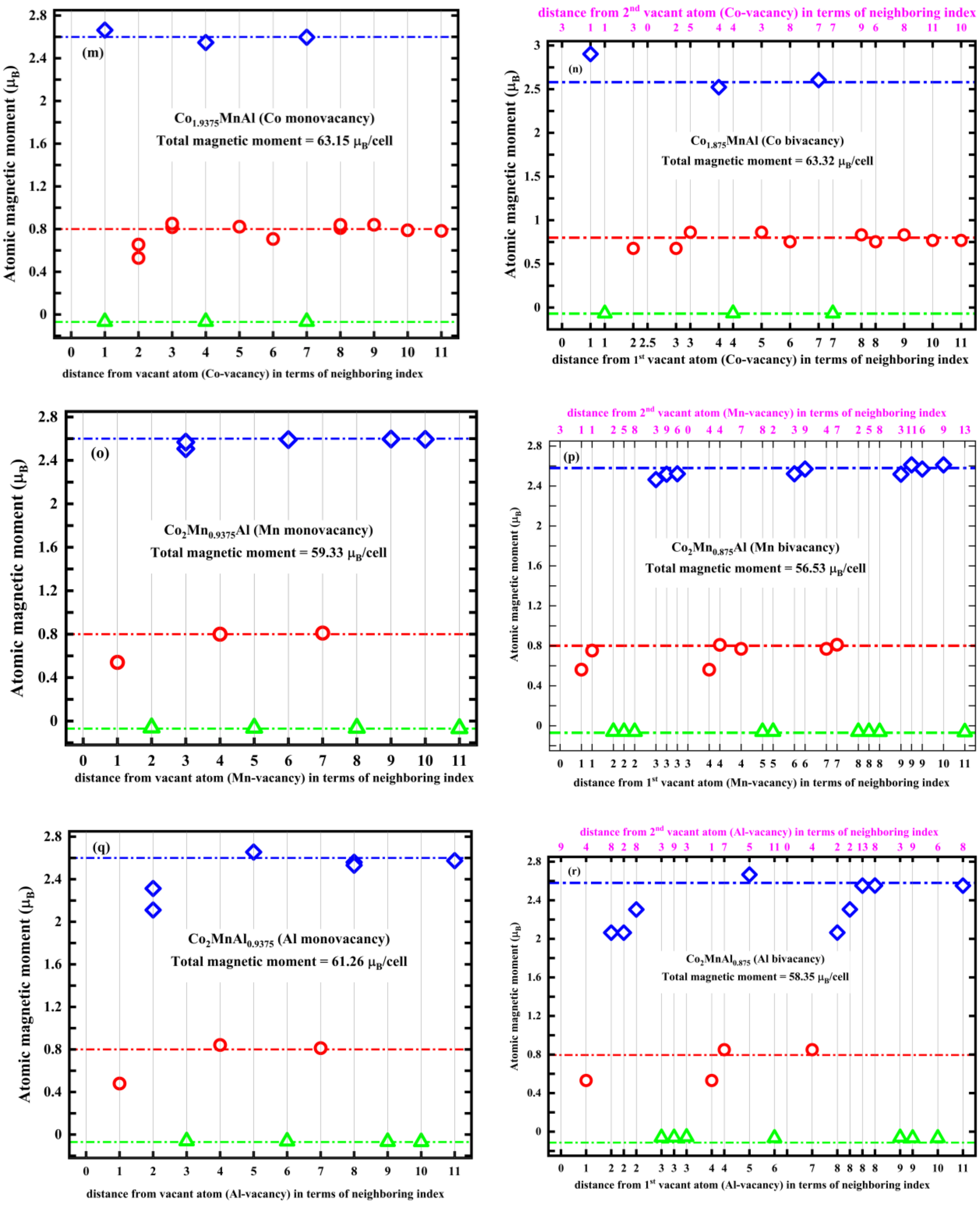

**Figure S2 :** Atomic magnetic moments for binary disordered structures and vacancy defected structures as a function of distance from the disordered atoms (or vacancy site for vacancy defected structures). The stoichiometric formula for each disordered structure, along with the total magnetic moment, is provided in the inset of every plot. In the case of disordered structures comprising two atoms, as depicted in Figures b, d, f, h, j, l, n, p, and r; the upper x-axis represents the distance from the second disordered atom or vacancy site.

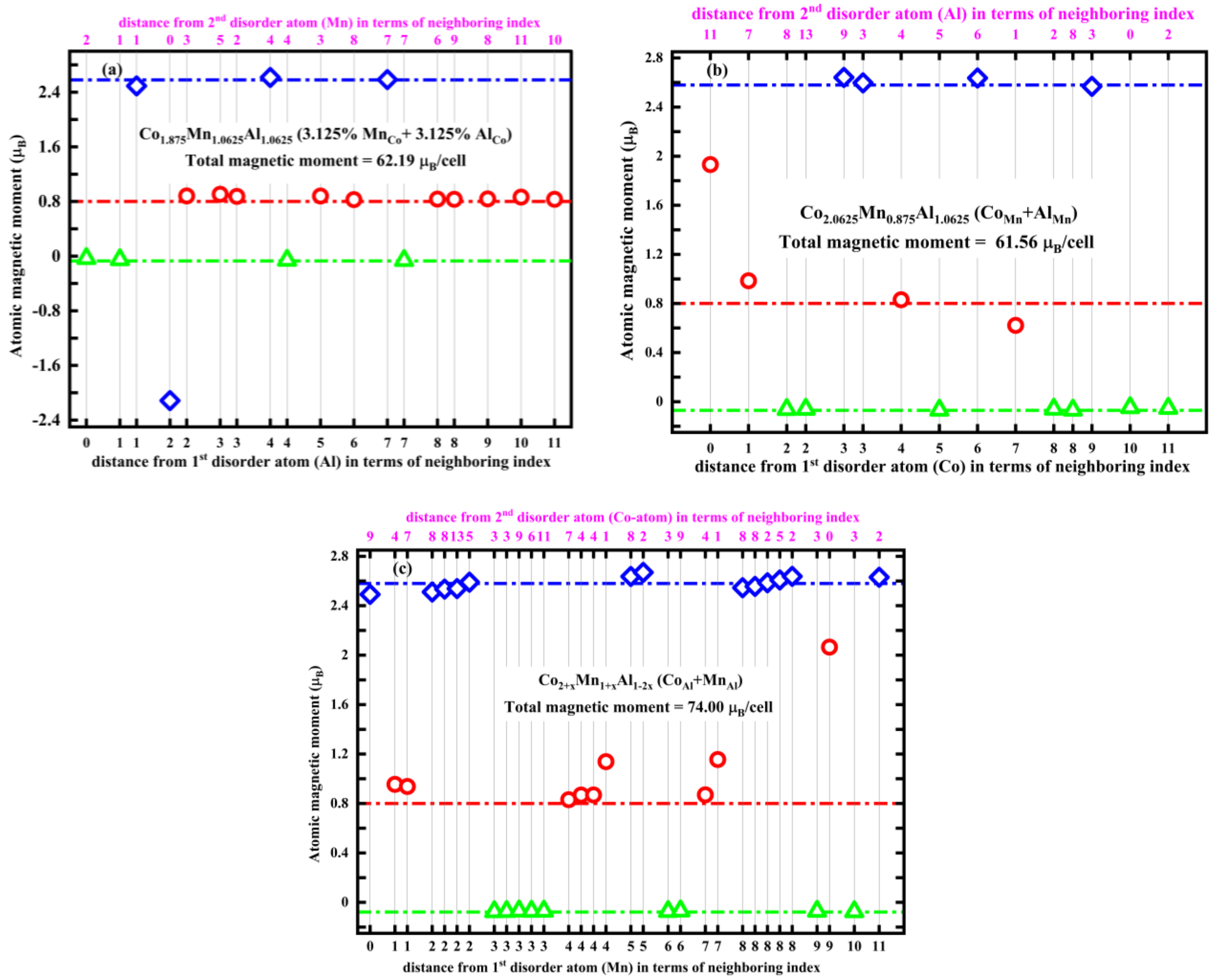

**Figure S3 :** (a), (b) and (c) present the AMM in similar way as in Figures S2, but for ternary antisite disordered structures.

## C. Structural, electronic, and magnetic properties for the uniform strained and tetragonal distorted structures

- **Lattice parameter(s), relative formation energy, spin polarization, total and atomic magnetic moment for the uniform strained structures:**

**Table S3:** The lattice parameter, relative formation energy (RFE), spin polarization, total and atomic magnetic moments for the uniform strained structures.

| **Percent change in volume $\Delta V/V_0$ (%)** | **Lattice parameter *a* (Å)** | **REF $E_{form}$(eV/*f.u.*)** | **Polarization *P* (%)** | **Total magnetic moment ($\mu_B$/*f.u.*)** | **Co-moment ($\mu_B$)** | **Mn-moment ($\mu_B$)** |
|---|---|---|---|---|---|---|
| -10.00 | 5.50 | 0.34 | 100.00 | 4.00 | 0.85 | 2.42 |
| -9.00 | 5.52 | 0.27 | 100.00 | 4.00 | 0.85 | 2.44 |
| -8.00 | 5.54 | 0.21 | 100.00 | 4.00 | 0.84 | 2.45 |
| -7.00 | 5.56 | 0.16 | 100.00 | 4.00 | 0.84 | 2.47 |
| -6.00 | 5.58 | 0.11 | 99.98 | 4.00 | 0.83 | 2.49 |
| -5.00 | 5.60 | 0.08 | 96.70 | 4.00 | 0.83 | 2.50 |
| -4.00 | 5.62 | 0.05 | 89.05 | 4.00 | 0.82 | 2.52 |
| -3.00 | 5.64 | 0.03 | 85.50 | 4.01 | 0.82 | 2.54 |
| -2.00 | 5.66 | 0.01 | 82.30 | 4.01 | 0.81 | 2.55 |
| -1.00 | 5.67 | 0.00 | 79.07 | 4.01 | 0.81 | 2.57 |
| 0.00 | 5.69 | 0.00 | 76.16 | 4.02 | 0.80 | 2.59 |
| 1.00 | 5.71 | 0.00 | 74.36 | 4.02 | 0.80 | 2.61 |
| 2.00 | 5.73 | 0.01 | 71.97 | 4.03 | 0.79 | 2.63 |
| 3.00 | 5.75 | 0.02 | 70.54 | 4.03 | 0.79 | 2.65 |
| 4.00 | 5.77 | 0.04 | 68.98 | 4.04 | 0.79 | 2.67 |
| 5.00 | 5.79 | 0.06 | 67.64 | 4.05 | 0.78 | 2.69 |
| 6.00 | 5.81 | 0.09 | 66.08 | 4.05 | 0.78 | 2.71 |
| 7.00 | 5.82 | 0.12 | 64.26 | 4.06 | 0.77 | 2.73 |
| 8.00 | 5.84 | 0.16 | 63.29 | 4.07 | 0.77 | 2.75 |
| 9.00 | 5.86 | 0.19 | 61.24 | 4.08 | 0.77 | 2.77 |
| 10.00 | 5.88 | 0.23 | 59.72 | 4.09 | 0.77 | 2.79 |

- **Lattice parameter(s), relative formation energy, spin polarization, total and atomic magnetic moments for the tetragonally distorted structures:**

**Table S4:** The lattice parameter(s), RFE, spin polarization, total and atomic magnetic moments for tetragonally distorted structure with V = ($V_0$-5%$V_0$) volume.

| c/a ratio | Lattice parameter *a* (Å) | Lattice parameter *c* (Å) | RFE ΔE (*eV/f.u.*) | Polarization *P (%)* | Total magnetic moment ($\mu_B$/*f.u.*) | Co-moment ($\mu_B$) | Mn-moment ($\mu_B$) |
|---|---|---|---|---|---|---|---|
| 0.50 | 7.05 | 3.53 | 12.53 | 29.73 | 1.54 | 0.05 | 1.15 |
| 0.60 | 6.64 | 3.98 | 5.40 | -1.53 | 1.18 | -0.11 | 1.21 |
| 0.70 | 6.30 | 4.41 | 2.49 | -66.92 | 2.74 | 0.69 | 1.31 |
| 0.80 | 6.03 | 4.82 | 0.97 | 89.39 | 3.94 | 0.84 | 2.32 |
| 0.85 | 5.91 | 5.02 | 0.51 | 97.74 | 4.01 | 0.85 | 2.40 |
| 0.90 | 5.80 | 5.22 | 0.24 | 97.23 | 4.00 | 0.84 | 2.45 |
| 0.95 | 5.69 | 5.41 | 0.12 | 95.15 | 4.00 | 0.83 | 2.48 |
| 1.00 | 5.60 | 5.60 | 0.08 | 96.70 | 4.00 | 0.83 | 2.50 |
| 1.05 | 5.51 | 5.78 | 0.11 | 96.97 | 4.00 | 0.83 | 2.49 |
| 1.10 | 5.42 | 5.96 | 0.17 | 97.26 | 4.00 | 0.83 | 2.48 |
| 1.15 | 5.34 | 6.14 | 0.29 | 95.43 | 4.00 | 0.83 | 2.45 |
| 1.20 | 5.27 | 6.32 | 0.41 | 48.88 | 3.53 | 0.64 | 2.32 |
| 1.30 | 5.13 | 6.67 | 0.49 | 31.71 | 2.28 | 0.11 | 2.05 |
| 1.40 | 5.00 | 7.00 | 0.63 | 23.13 | 2.07 | 0.05 | 1.95 |
| 1.50 | 4.89 | 7.33 | 0.88 | 6.04 | 1.96 | 0.03 | 1.86 |

**Table S5:** Same as Table S4, but with $V = V_0$ volume.

| c/a ratio | Lattice parameter *a* (Å) | Lattice parameter *c* (Å) | RFE ΔE (*eV/f.u.*) | Polarization *P (%)* | Total magnetic moment ($\mu_B$/*f.u.*) | Co-moment ($\mu_B$) | Mn-moment ($\mu_B$) |
|---|---|---|---|---|---|---|---|
| 0.50 | 7.17 | 3.59 | 11.4 | 16.66 | 1.62 | 0.04 | 1.25 |
| 0.60 | 6.75 | 4.05 | 4.9 | -12.61 | 1.13 | -0.23 | 1.41 |
| 0.70 | 6.41 | 4.49 | 2.26 | -50.72 | 3.04 | 0.75 | 1.52 |
| 0.80 | 6.13 | 4.91 | 0.83 | 51.66 | 4.05 | 0.86 | 2.44 |
| 0.85 | 6.01 | 5.11 | 0.4 | 87.57 | 4.03 | 0.84 | 2.49 |
| 0.90 | 5.90 | 5.31 | 0.16 | 85.72 | 4.02 | 0.82 | 2.54 |
| 0.95 | 5.79 | 5.50 | 0.03 | 79.53 | 4.02 | 0.81 | 2.57 |
| 1.00 | 5.69 | 5.69 | 0 | 76.16 | 4.02 | 0.8 | 2.59 |
| 1.05 | 5.60 | 5.88 | 0.02 | 78.16 | 4.02 | 0.8 | 2.58 |
| 1.10 | 5.51 | 6.07 | 0.09 | 80.65 | 4.02 | 0.81 | 2.57 |
| 1.15 | 5.43 | 6.25 | 0.19 | 54.65 | 4.05 | 0.82 | 2.55 |
| 1.20 | 5.36 | 6.43 | 0.32 | 42.55 | 4.19 | 0.9 | 2.53 |
| 1.30 | 5.22 | 6.78 | 0.45 | 41.72 | 2.36 | 0.07 | 2.23 |
| 1.40 | 5.09 | 7.12 | 0.59 | 36.62 | 2.1 | -0.05 | 2.15 |
| 1.50 | 4.97 | 7.46 | 0.83 | 42.03 | 2.05 | 0.12 | 2.1 |

**Table S6:** Same as Table S4, but with V= ($V_0$ +5%$V_0$) volume.

| c/a ratio | Lattice parameter *a* (Å) | Lattice parameter *c* (Å) | RFE ΔE (*eV/f.u.*) | Polarization *P (%)* | Total magnetic moment ($\mu_B$/*f.u.*) | Co-moment ($\mu_B$) | Mn-moment ($\mu_B$) |
|---|---|---|---|---|---|---|---|
| 0.50 | 7.29 | 3.65 | 10.51 | -3.52 | 1.72 | 0.03 | 1.36 |
| 0.60 | 6.86 | 4.12 | 4.64 | -37.92 | 3.82 | 0.94 | 1.78 |
| 0.70 | 6.52 | 4.56 | 2.12 | -33.07 | 5.48 | 1.42 | 2.60 |
| 0.80 | 6.23 | 4.99 | 0.81 | 16.37 | 4.61 | 1.10 | 2.61 |
| 0.85 | 6.11 | 5.19 | 0.43 | 38.30 | 4.14 | 0.84 | 2.61 |
| 0.90 | 5.99 | 5.39 | 0.20 | 74.00 | 4.07 | 0.81 | 2.64 |
| 0.95 | 5.89 | 5.59 | 0.10 | 70.03 | 4.05 | 0.79 | 2.67 |
| 1.00 | 5.79 | 5.79 | 0.06 | 67.64 | 4.05 | 0.79 | 2.69 |
| 1.05 | 5.69 | 5.98 | 0.09 | 67.41 | 4.05 | 0.79 | 2.68 |
| 1.10 | 5.61 | 6.17 | 0.15 | 55.90 | 4.06 | 0.79 | 2.66 |
| 1.15 | 5.52 | 6.35 | 0.25 | 47.98 | 4.19 | 0.86 | 2.66 |
| 1.20 | 5.45 | 6.53 | 0.35 | 41.51 | 4.35 | 0.93 | 2.65 |
| 1.30 | 5.30 | 6.89 | 0.54 | 47.20 | 2.58 | 0.08 | 2.44 |
| 1.40 | 5.17 | 7.24 | 0.76 | 7.26 | 4.42 | 1.02 | 2.52 |
| 1.50 | 5.05 | 7.58 | 0.93 | -27.51 | 5.08 | 1.28 | 2.64 |

- **Fermi energies for the uniform strained and tetragonal distorted structures:**

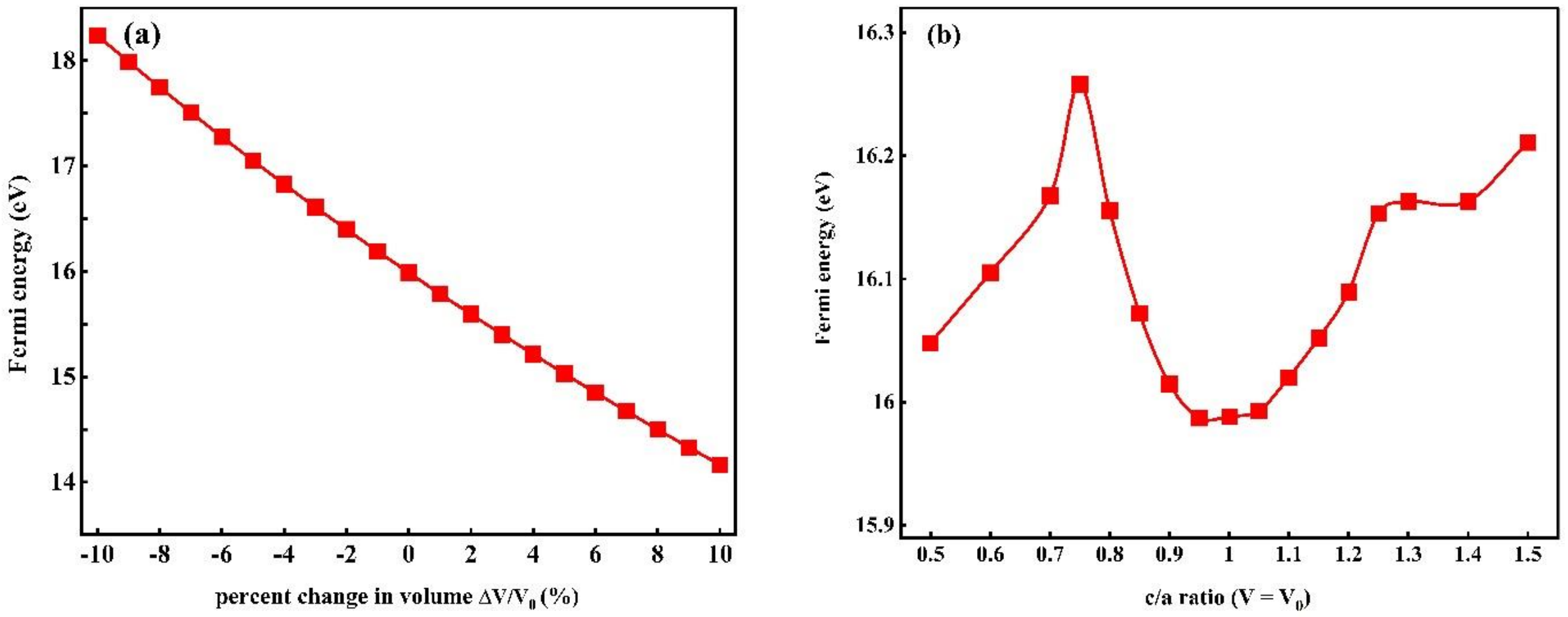

**Figure S4**: Variation in the Fermi energy for the (a) uniform cubic strained structure as a function of percent change in the unit cell volume $\Delta V_0/V$ and (b) one of the tetragonally distorted structures (with $V_0$ volume) as a function of c/a ratio.

- **Partial DOS for uniform strained structures:**

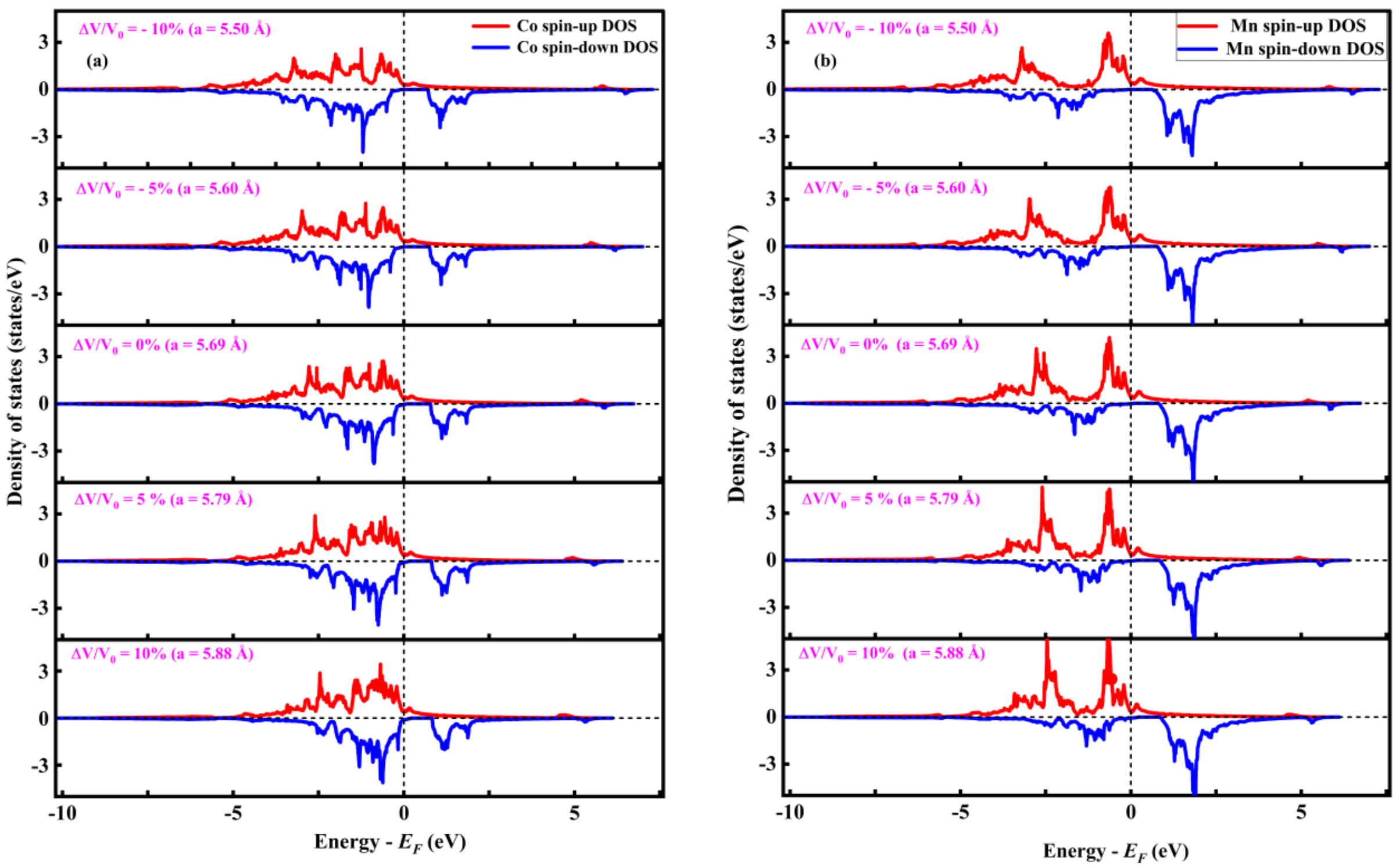

**Figure S5**: Partial DOS for (a) Co atoms and (b) Mn atoms for 0, ±5, ±10% (= $\Delta V_0/V_0$) uniform strained structures. The red and blue solid lines show the majority (spin up) and minority (spin down) DOS.